\documentclass[sigconf,nonacm]{acmart}
\AtBeginDocument{%
  }

\usepackage{caption}
\usepackage{subcaption}
\usepackage{todonotes}
\usepackage{pifont}
\usepackage{xspace}
\usepackage{CJKutf8}
\usepackage{makecell}
\usepackage{enumitem}
\usepackage{listings}
\usepackage{amsmath}
\usepackage{textcomp}
\usepackage{tabularx}
\usepackage{booktabs}
\usepackage{xcolor} 
\usepackage{fvextra}

\usepackage{longtable}
\usepackage{booktabs}
\usepackage{array}

\newcommand{\ie}{{\em i.e.,\/ }}
\newcommand{\eg}{{\em e.g.,\/ }}
\newcommand{\vs}{{\em vs.\/ }}

\newcommand{\pb}[1]{\vspace{0.5ex}\noindent{\bf \em #1}\hspace*{.3em}}

\newcommand{\one}{({\em i}\/)\xspace}
\newcommand{\two}{({\em ii}\/)\xspace}
\newcommand{\three}{({\em iii}\/)\xspace}

\newcolumntype{L}[1]{>{\raggedright\let\newline\\\arraybackslash\hspace{0pt}}m{#1}}
\newcolumntype{C}[1]{>{\centering\let\newline\\\arraybackslash\hspace{0pt}}m{#1}}
\newcolumntype{R}[1]{>{\raggedleft\let\newline\\\arraybackslash\hspace{0pt}}m{#1}}

\newcommand{\zh}[1]{\begin{CJK}{UTF8}{gbsn}#1\end{CJK}}

\begin{document}

\title{Navigating the Open-Source Model Ecosystem: An Empirical Study of Creator Practices in Artistic Image Generation}



\settopmatter{authorsperrow=4}

\author{Yiluo Wei}
\affiliation{%
  \institution{The Hong Kong University of Science and Technology (Guangzhou)}
  \city{Guangzhou}
  \country{China}
}

\author{Yupeng He}
\affiliation{%
  \institution{The Hong Kong University of Science and Technology (Guangzhou)}
  \city{Guangzhou}
  \country{China}
}

\author{Qiming Ye}
\affiliation{%
  \institution{The Hong Kong University of Science and Technology (Guangzhou)}
  \city{Guangzhou}
  \country{China}
}

\author{Gareth Tyson}
\affiliation{%
  \institution{The Hong Kong University of Science and Technology (Guangzhou)}
  \city{Guangzhou}
  \country{China}
}







\begin{abstract}
  The open-sourcing of powerful image generation models has created a vibrant ecosystem where creators curate and combine a vast array of community-contributed models. This practice stands in sharp contrast to using closed-source tools like Midjourney. Yet, little is known about these emerging creative workflows. To bridge this gap, this paper presents the first large-scale empirical study of creator model usage behavior within this open-source image generation ecosystem. We construct a novel dataset of 6 million images with their embedded generation metadata --- a detailed recipe of the creation process, including the models used and the prompts. By linking the usage of 22.4K base models and 154K LoRA models to the images, our findings underscore the ecosystem's unique strengths and its inherent obstacles. This provides valuable insights for making this ecosystem more sustainable and innovative. Moreover, we make our dataset publicly available,\footnote{\url{https://huggingface.co/datasets/Wei-Yiluo/Pixiv-AI-generation-metadata}} providing creators with practical references for producing better artworks and researchers to facilitate further studies.
\end{abstract}




\maketitle

\section{Introduction}

The commodification of AI Generated Content (AIGC) has reshaped online creative communities. The advent of powerful generative diffusion models \cite{diffusion} has been a key catalyst in this transformation \cite{dhariwal2021diffusion}. The open-sourcing of highly capable models like Stable Diffusion \cite{stable-diffusion} has further democratized access to this technology, allowing users not only to generate images but also to easily fine-tune and extend models for specific styles and concepts \cite{10204880, 10377881, gal2023an}.

This accessibility has fueled the rapid emergence of a vibrant open-source generative AI ecosystem \cite{icml24near}. Platforms like Civitai \cite{MM24b} now host hundreds of thousands of community-contributed models. In this new paradigm, different from the close-source commercial platforms such as Midjourney \cite{midjourney, 10.1145/3626235}, creators are no longer passive users of a single tool. Instead, they are active curators who can select from, combine, and even develop a diverse array of models to realize their artistic visions \cite{10.1016/j.infoandorg.2025.100560}.  
For example, a common practice is to combine a foundational ``base'' model with lightweight LoRA (Low-Rank Adaptation) finetuning \cite{lora} for nuanced adjustments.
This complex interplay of tools, techniques, and community-driven development represents a fundamental shift in digital creation.

However, while the technical capabilities and the proliferation of models have been well documented \cite{10081412, 10419041}, a critical gap remains in our understanding of the creative practices that have emerged within this ecosystem --- specifically, how this massive and diverse array of models are selected, used, and combined by creators.
To fully appreciate the value of the open-source paradigm, it is crucial to systematically investigate the key differences in its highly modular creative workflows compared to close-source alternatives.
Our goal is to identify its unique strengths and inherent challenges. Such an understanding will directly benefit creators seeking to master new workflows to empower their artworks, developers aiming to build more useful models and tools, and platforms striving to foster a sustainable and innovative community \cite{doi:10.1126/science.adh4451, Fathoni2023}. Thus, to bridge this gap, wee present the first large-scale empirical study into creators' model usage behavior within this open-source ecosystem.

Our goal is different to prior efforts in this space. 
Previous research has often relied on large-scale prompt-image datasets such as DiffusionDB \cite{wang2022diffusiondb, sun2023journeydb} to study creator behavior \cite{3491102.3501825, textbaseimage_survey, Oppenlaender17112024}. However,  their focus is the prompting behavior while interacting with a single generative model.
In contrast, this study focuses a different scenario: investigating model usage behavior within an ecosystem comprising hundreds of thousands of models.
There are also available datasets from model-sharing platforms like Civitai \cite{Civiverse}. However, these platforms are primarily repositories for model distribution, and the images hosted there often serve as technical previews or examples to showcase a model’s range. While these datasets may accurately represent the model developers, they do not represent the creator community required for this study.



To overcome previous limitations and capture authentic creative practices, we conduct an empirical case study on Pixiv to examine how creators utilize AI models in real-world artistic endeavors. As one of the world's largest online art communities, hosting over 140 million artworks and 1 billion monthly views \cite{MM24a}, Pixiv remains open to AIGC while enforcing strict disclosure policies. This transparency makes it an ideal, well-documented environment for studying AIGC production \cite{MM24a, Kwon2024, kim2024generative}.

To conduct this study, we constructed a novel, large-scale dataset by systematically collecting AI-tagged images on Pixiv and extracting the generation metadata embedded within their file headers. This metadata provides the exact ``recipe'' of the artwork, including prompts, generation parameters, and specific models used. We linked these configurations to platform engagement metrics (\eg views, likes) and cross-referenced the extracted models with major model-sharing platforms to determine their characteristics. Overall, our dataset comprises 6 million images with complete generation metadata, spanning 22.4K base models and 154K LoRA models.
We make this curated dataset publicly available. This resource will provide creators with practical references for generating better artworks, while enabling future research in model recommendation \cite{gemrec}, prompt auto-completion \cite{chatgen, diffagent}, and the socio-technical dynamics of AIGC~\cite{MM24a}.
Using this dataset, we investigate the following research questions:
\begin{itemize}[leftmargin=*]

    \item \textbf{RQ1: What is the distribution and variety of model usage within the open-source ecosystem?}
    A key difference between the open-source paradigm and the closed-source tools is the vast selection of models. We thus examine the distribution of model usage and the extent to which individual creators leverage a variety of models.

    \item \textbf{RQ2: What are the life cycles of models?} 
    This vibrant ecosystem is in constant flux, with new models emerging continuously. Therefore, to understand its evolution, we analyze the temporal dynamics of model usage, \ie how a model begins to gain popularity, reaches its peak usage, and eventually becomes obsolete.

    \item \textbf{RQ3: How do creators utilize LoRA models for fine-grained customization?} 
    The capability to integrate a base model with  LoRA adaptations is a distinctive feature of this ecosystem. We explore how creators employ and combine LoRAs, identifying the evolution of patterns and strategies and their effectiveness.

\end{itemize}
\vspace{-0.5ex}
Our analysis reveals several key findings: 
\vspace{-0.5ex}
\begin{enumerate}[leftmargin=*]
    \item The ecosystem is vast, with over 22.4K unique base models and 154K LoRA models. However, this diversity is not fully realized in practice. We find that model usage follows a classic long-tail distribution, where a small fraction of popular models dominates image generation, with 80\% of images generated by only 560 (2.5\%) base models. Further, the majority of creators rely on a very limited set of familiar models, and only the highly prolific, expert users experiment with larger number of different models. These observations suggest that the sheer volume of options may overwhelm creators, leading them to gravitate towards a small set of popular choices. This points to a need for better support systems, such as model recommendation engines, to widen access to the full creative potential of the open-source landscape for creators at all skill levels. (\S\ref{sec:rq1})
    
    \item Models tend to have long life cycles, but often experience slow adoption within the community. Base models follow a ``rise and fall'' pattern, gaining popularity quicker, with 50\% reaching peak usage in 11 weeks. LoRA models are adopted more slowly but often maintain their utility, with 50\% reaching peak usage in 23 weeks. This persistence demonstrates a unique strength of the open ecosystem, where a vast collection of tools is maintained to serve the long-tail requirements of sub-cultures and personalized artistic expression. 
    However, the slow community adoption also emphasizes the need for intelligent recommendation systems and automated workflow adaptation tools to better assist creators discovering and integrating the most advanced and suitable models. 
    (\S\ref{subsec:rq2:temporal_popularity})
    
    \item One key reason why models have long life cycles is the inertia creators exhibit when it comes to updating their model versions. On average, only 40\% of images adopt the latest version even after 20 weeks of its release. While partially (\textasciitilde40\%) explained by a natural delay in adoption, this behavior can also be attributed to a deliberate choice by artists who favor the specific aesthetic qualities of a particular version. This highlights a core value of the open ecosystem: unlike commercial platforms that can force updates, users here retain the autonomy to choose their preferred tools. (\S\ref{subsec:rq2:version})

    \item There has been a dramatic surge in LoRA adoption, with 75\% of all images utilizing LoRAs in 2025. The use of LoRA is highly correlated with artwork popularity, with LoRA-enhanced images receiving more views and bookmarks. As a unique feature of this open-source ecosystem, these results confirm LoRA’s importance and its effectiveness in the creative process. (\S\ref{subsec:rq3:lora_adoption})

    \item The use of LoRA is becoming increasingly sophisticated. Basic functional tasks, such as drawing a specific character, are offloaded to more capable or specialized base models, which allows the capacity of LoRA to be redirected towards stylistic and conceptual customization or more nuanced adjustments. This shift indicates a more advanced and specialized creative process. However, the increasing complexity may pose a challenge for new or amateur creators, underscoring the need for future tools to assist users in navigating the use of LoRAs, including their compatibility and potential conflicts. (\S\ref{subsec:rq3:lora_category})
    
\end{enumerate}

\begin{figure*}
    \centering
    \includegraphics[width=\textwidth]{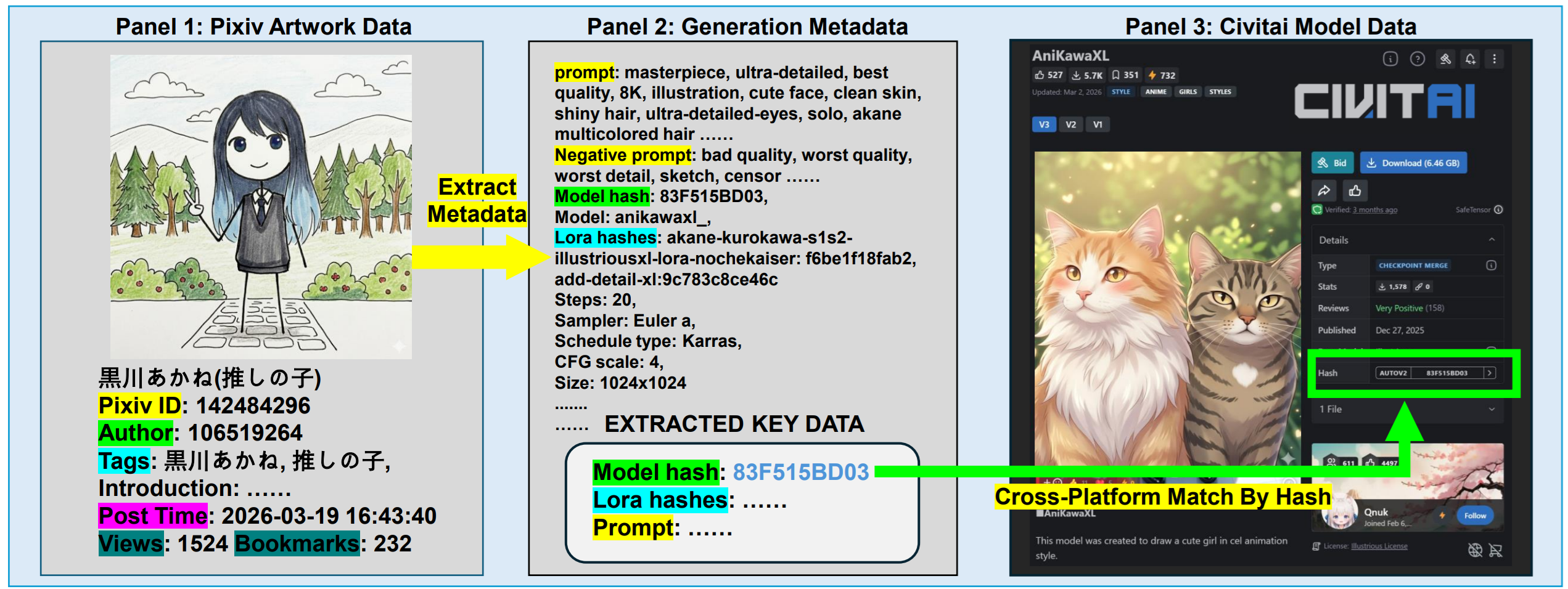}
    \vspace{-5.5ex}
    \caption{Overview of the method for data collection and processing.}
    \vspace{-1.3ex}
    \label{fig:datacollection}
\end{figure*}

\section{Related Work}
\pb{Prompt-Image Datasets \& Their Usage.} 
The rapid adoption of Text-to-Image (T2I) models~\cite{textbaseimage_survey} has led to the creation of large-scale prompt-image datasets to study user behavior.
Early efforts such as DiffusionDB~\cite{wang2022diffusiondb} collected millions of prompt-image pairs from Discord to analyze prompt structure and keyword usage.
Similarly, JourneyDB~\cite{sun2023journeydb} focused on high-quality Midjourney~\cite{midjourney} data to support multimodal analysis.
To capture human preferences, Pick-a-Pic~\cite{kirstain2023pick} gathered prompt-image interactions through a Discord bot, enabling research on preference alignment.
Beyond analysis, these datasets also support practical tasks, such as preference alignment in ImageGem~\cite{imagegem} and detection benchmarking in COCOXGEN~\cite{cocoxgen}.
However, for these works, the underlying generative model is typically a monolithic entity, which is unable to capture the diverse selection and combination of specific models that define the contemporary open-source ecosystem. 
The most similar dataset to ours is Civiverse \cite{Civiverse}, compiled from images on Civitai. However, Civitai primarily serves as a platform for model sharing rather than artistic creation, reflecting different use cases.

\pb{The Open-source Gen-AI Ecosystem for Image Creation.} 
The open-source generative AI ecosystem has rapidly expanded \cite{MM24a, MM24b} with the growth of base models and parameter-efficient fine-tuning methods such as LoRA~\cite{lora, cloneofsimo_lora}. 
This expansion makes it increasingly challenging to evaluate and select models, since the large number of variants and fine-tuned versions complicates direct comparison and benchmarking~\cite{xu2024benchmarkingbenchmarkleakagelarge,ramamoorthy2025evaluating,joonbakhsh2025evidence}.
Platforms like GenAI Arena~\cite{genai_arena} address this by ranking models using crowdsourced human preferences.
At the same time, the large number of customized models makes selection a bottleneck.
To address this, DiffAgent~\cite{diffagent} and ChatGen~\cite{chatgen} use LLM-based agents to automatically select suitable models and parameters from user queries.
Other work frames this as a recommendation task, such as GEMRec~\cite{gemrec}, which suggests models based on prompts.
To handle unreliable metadata, CARLoS~\cite{carlos} proposes content-based retrieval that selects suitable LoRAs based on their visual effects rather than text descriptions.
While these efforts are valuable, they often focus on specific subtasks (\eg recommendation) under controlled conditions.
In contrast, our goal is to evaluate the creators' model usage behavior in real-world scenarios.

\section{Data Collection and Processing}
\label{sec:data}

\subsection{Pixiv Artwork Data}
\label{subsec:data:pixiv}
Pixiv is a Japanese platform and online community for artists, founded in Tokyo in 2007. By 2026, it hosts over 130 million artistic submissions. The site garners more than 600 million page views monthly \cite{pixiv-stat}. Pixiv's main aim is to provide artists with a space to display their illustrations and receive feedback through comments. The platform primarily features original artwork inspired by Japanese anime and manga, while generally excluding photography.

Pixiv was a pioneer in allowing the upload of AI-generated images and with new policies introduced in late October 2022 \cite{pixiv-ai-policy}:
\one Artists are required to explicitly indicate whether their submissions are human-generated or AI-generated using a special toggle. This information is displayed when others view the submission.
\two Users have the option to filter out all AI-generated content if they prefer.
\three AI-generated and human-generated submissions are ranked separately.
These policies establish Pixiv as a leading platform for sharing AI-generated works \cite{MM24a}, making it an ideal case study for examining the usage of AI models.

\pb{Collecting Pixiv Artworks.}
Creators submit their images to Pixiv as \emph{Artworks}, which is similar to a post on Instagram. 
An artwork on Pixiv can encompass multiple images. 
Each artwork also contains a suite of metadata, of which, the creator, the timestamp, the tags (an array of up to 10 tags added by the creator or the viewers), and cumulative view and bookmark counts are used in this study.
We collect all the AI-tagged artworks from \texttt{2023-01-01} to \texttt{2025-12-31} with the metadata described above.

\pb{Extracting Generation Metadata from Images.}
Then we check the images in each artwork and only include the artwork if we can get complete  ``generation metadata'' from the images, which contains the full configuration of the models used, the hyperparameters set, and the prompt.
For each image, we extract and process the generation metadata. Among them, the following are used in our analysis: \one The base/foundational model used (identified by a hash value); \two the LoRA models used (identified by a hash value); \three The prompt and negative prompt. 
Overall, we get 860,075 artworks from 18,430 creators, with 5,977,372 images.

\subsection{Civitai Model Data}
\label{subsec:data:model}

Civitai is a prominent community-driven repository and platform dedicated to open-source generative artificial intelligence models. It hosts a diverse array of community-contributed models, primarily built upon the Stable Diffusion architecture for image generation. These include full-size base diffusion models (\ie ``checkpoints''),  lightweight LoRA finetuning models, and other related models that may play a role in the image generation process.
At the time of writing, Civitai hosts more than 500K models \cite{civitai_changelog}.
By a significant margin, it has become the largest and most widely utilized platform of its kind, serving as the primary data source for both artists and researchers \cite{MM24b, carlos, diffagent, 3715275.3732158, ghosh2026marketplace, 3715275.3732107, Civiverse}.

\pb{Collecting Model Metadata.}
Models on Civitai are organized into \emph{model families}, each comprising multiple \emph{model versions} that typically result from version updates (\ie v1 -> v2 -> v3). For clarity, throughout this paper, \textit{the term ``model'' refers to a specific version within a model family}, which is identified by a unique hash value. Each model comes with a set of metadata, of which, the unique model family and version ID, publication date, and model category are utilized in this study. We collect the metadata for all models available on Civitai by \texttt{2025-12-31}, totaling 387,242 models.

\pb{Model Cross-Reference.}
The image generation metadata, as described in \S\ref{subsec:data:pixiv}, includes the hash value of models. We utilize this hash value to reliably connect the model identified in the generation metadata to its counterpart on Civitai. This linkage allows us to perform analysis using the model metadata from Civitai effectively.

\section{RQ1: Model Diversity}
\label{sec:rq1}

A fundamental distinction of the open-source ecosystem from  closed-source platforms is the vast and diverse array of independent models available to creators. To understand the practical implications of this choice, we begin by addressing RQ1, which quantifies the diversity of model usage.

\subsection{Overview of the Number of Models}    
\label{subsec:rq1:overview}

\begin{figure}[t]
    \centering
    \includegraphics[width=\linewidth]{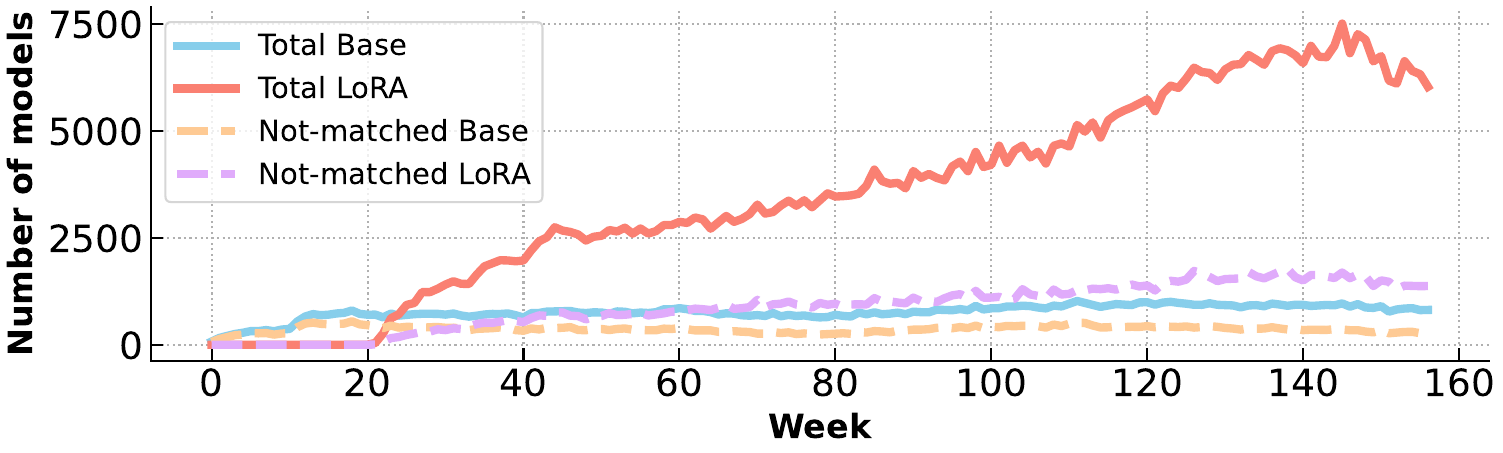}
    \vspace{-5.3ex}
    \caption{Weekly count of unique models used.}
    \vspace{-3ex}
    \label{fig:rq1:num_of_models}
\end{figure}

We begin by presenting the number of models identified in our dataset. 
Across 5,977,372  million images, we identified 22,412 unique base models, and 153,973 LoRA models. This suggests a remarkably large and diverse ecosystem. 
To understand the temporal dynamics, we then calculate the per-week results which are plotted in Figure \ref{fig:rq1:num_of_models}.
We see a sustained growth in the number of unique LoRA models adopted each week, peaking at over 7,500 and finally see a slight drop at the end of our observation period. This highlights the creative exploration and diversification with the rapid proliferation of lightweight, specialized LoRAs (we further study the use of LoRAs in details in \S\ref{sec:rq3}). 

This is in contrast to the trend for base models. While the number of unique base models saw an initial increase in the first 15 weeks, it quickly stabilized, fluctuating around a relatively constant $\sim$700 unique models per week. 
However, we also note the significant discrepancy between the $\sim$700 models used in a given week and the 22,412 unique base models we identified in total. This probably indicates that the base models being used is in constant flux, with new and improved base models continuously entering the ecosystem and competing with older ones, which prompts us to further study the life cycle of the models in \S\ref{sec:rq2}.

Additionally, the figure also reveals a persistent use of ``not-matched'' models (weekly 47.5\% of base models and 25.4\% of LoRA models).
These models are not found on Civitai, suggesting a widespread practice of creators training their own bespoke models or sharing them within smaller, private communities. This finding further substantiates the idea that creators are not merely passive consumers of publicly available tools but are also active producers and customizers, tailoring the technology to their specific creative needs.
However, the high prevalence of unshared models also points to a concern regarding ecosystem fragmentation: while it is understandable that some of models cannot be publicly shared for a good reason, if a large portion of innovation remains siloed in private repositories, the community loses the benefits of collective refinement. To address this issue, the ecosystem may need to reduce the friction (\eg integrating the creation process with public model repositories) and introduce incentives for sharing.

\subsection{Model Usage Distribution across Images}
\label{subsec:rq1:distribution}

In \S\ref{subsec:rq1:overview}, we observe a large number of models employed in image creation. However, if model adoption is concentrated on just a few models, we cannot confidently conclude that there is good diversity. Hence, we examine the distribution of model usage per image.

\begin{figure}[t]
    \centering
    \includegraphics[width=.49\linewidth]{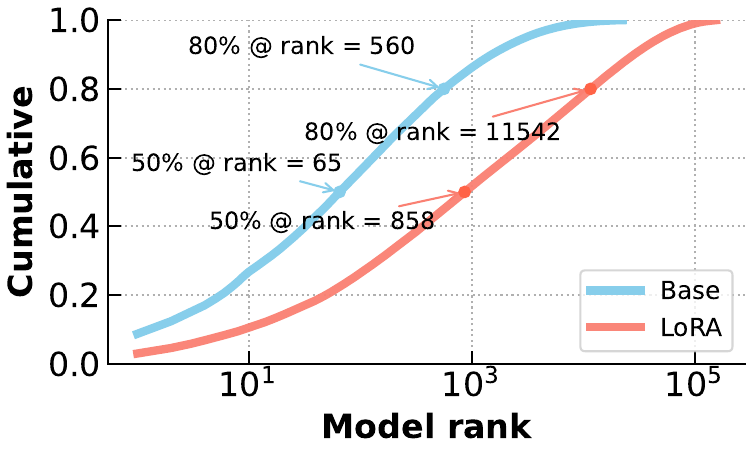}
    \includegraphics[width=.49\linewidth]{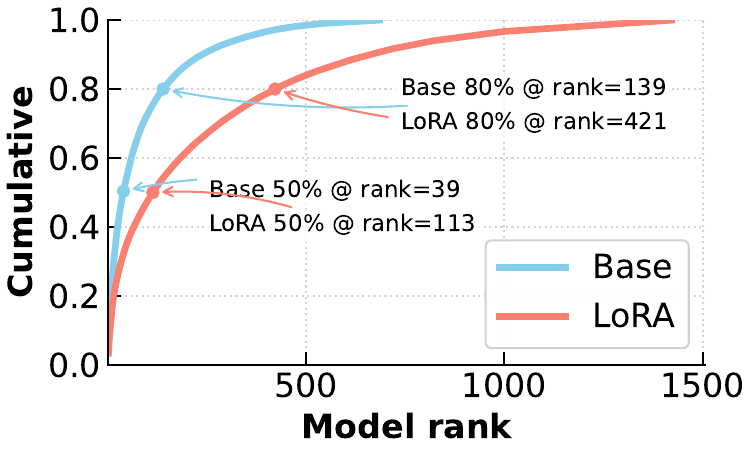}
    \vspace{-2.5ex}
    \caption{Adoption distribution of base and LoRA models: (a) Over three years; (b) The week with largest entropy, where base model is \texttt{2023-08-08} and LoRA model is \texttt{2023-08-15}.}
    \vspace{-1ex}
    \label{fig:CDFofmodel_rank_and_firstpeak}
\end{figure}

\pb{Overall Result.}
Figure \ref{fig:CDFofmodel_rank_and_firstpeak}a presents the adoption distribution of both base and LoRA models over the whole three years in our dataset.
In this figure, usage is dominated by a relatively small set of popular models, revealing a classic long-tail distribution. For base models, despite 22,412 unique options being available, the top 65 models alone account for 50\% of all generated images, and the top 560 models cover 80\% of usage. For LoRA models, despite 153,973 unique options being available, the top 858 models alone account for 50\% of all generated images, and the top 11,542 models cover 80\% of usage.
This heavy concentration indicates that the diversity of model usage is not as pronounced as the large number of unique models suggests, with the vast majority of models existing in a long tail of niche or infrequent use.
However, this finding should not be simply interpreted as a lack of diversity. In fact, the ``head'' of this distribution still comprises hundreds base and thousands of LoRA models. While creators may gravitate toward a preferred set of tools, that set is orders of magnitude larger and more varied than any closed-source alternative.

\begin{figure}[t]
    \centering
    \includegraphics[width=\linewidth]{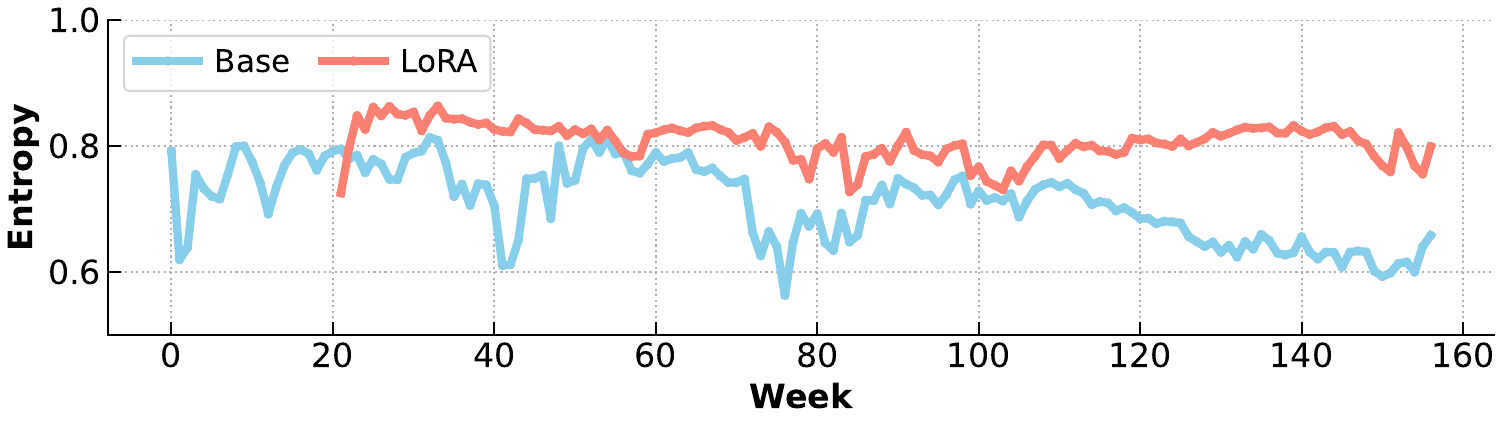}
    \vspace{-5.2ex}
    \caption{Weekly Shannon entropy of model adoption.}
    \vspace{-3ex}
    \label{fig:rq1:entropy}
\end{figure}

\pb{Weekly Results.}
While the overall distribution provides a valuable macro-level view, aggregating three years of data into a static snapshot may obscure the rapid turnover of models we established earlier. That is, although only $\sim$700 base models are actively used in a given week, the three-year data aggregates all 22,412 models.
To address this, we shift our analysis to a weekly level to observe how the diversity landscape fluctuates over time.

Figure \ref{fig:CDFofmodel_rank_and_firstpeak}b illustrates the model adoption distribution during the week with the highest normalized Shannon entropy \cite{cover1999elements}, \ie the most ``diverse'' week. 
During this period, 139 ($\sim$20\%) base models are required to capture 80\% of the cumulative usage. 
Notably, if it is over the entire three-year span (Figure \ref{fig:CDFofmodel_rank_and_firstpeak}a), the top 2.5\% of models generate 80\% of the images. This significant difference in proportions (20\% \vs 2.5\%) demonstrates that usage concentration is much less extreme at the weekly level than the long-term aggregate implies. The same pattern is observed for LoRA models.

This suggests that the system's short-term dynamics are likely governed by a stable ``head'' of a few hundred routinely used models that evolve slowly over time (\ie the set of weekly popular models changes gradually). Meanwhile, the remaining models turn over rapidly from week to week, accumulating into a massive, highly dynamic long tail over the years. We further explore this in \S\ref{sec:rq2}.

As the weekly count of unique models changes (Figure~\ref{fig:rq1:num_of_models}), we also assess the diversity of the model ecosystem over time. Figure~\ref{fig:rq1:entropy} illustrates the normalized Shannon entropy of model adoption over the three-year period.
We see that the number of adopted LoRA models grows continuously week over week, while maintaining a consistently high entropy (hovering near 0.8).
In contrast, the entropy of base models exhibits a steady downward trend during the recent year (2025), declining from a peak of approximately 0.8 to a low of 0.6. This suggests a concerning trend where a few popular base models are increasingly dominating user preferences.

\subsection{Creators' Model Usage Patterns}

In the previous subsection (\S\ref{subsec:rq1:distribution}), we found that although many models are available, only a small subset is heavily used. To further understand this pattern, this subsection shifts the focus to individual creators' behavior, investigating whether creators tend to use a diverse range of models. 

\begin{figure}[h!]
    \centering
    \includegraphics[width=.49\linewidth]{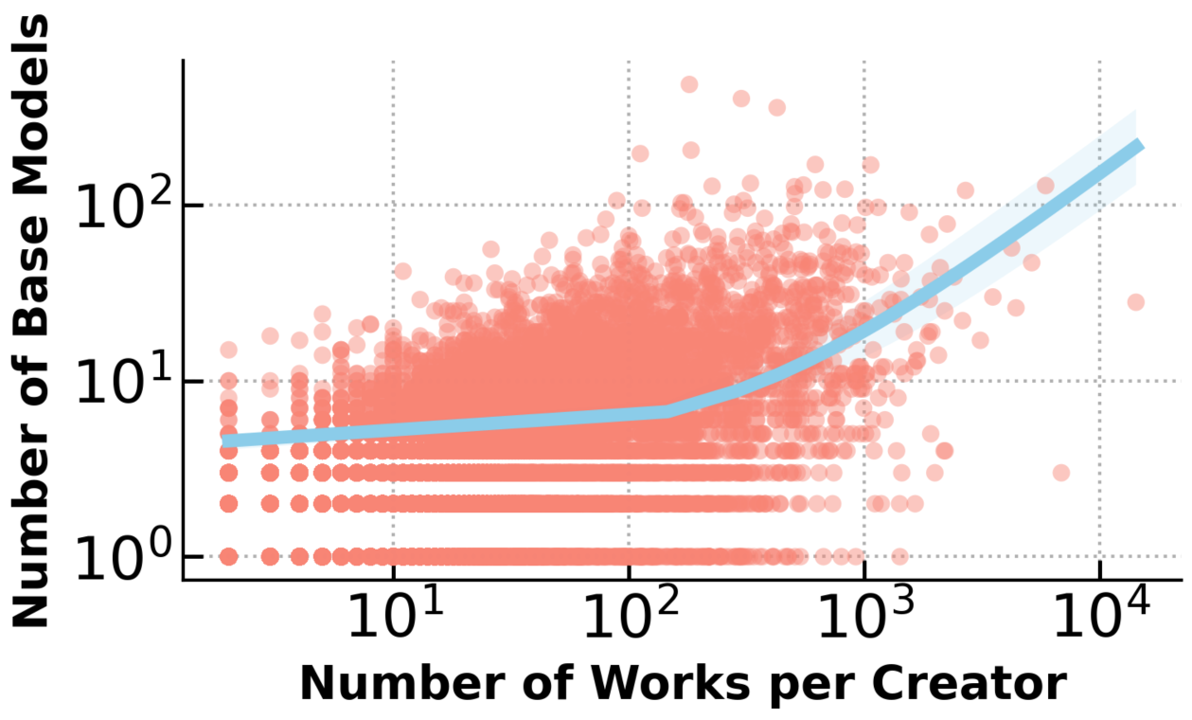}
    \includegraphics[width=.49\linewidth]{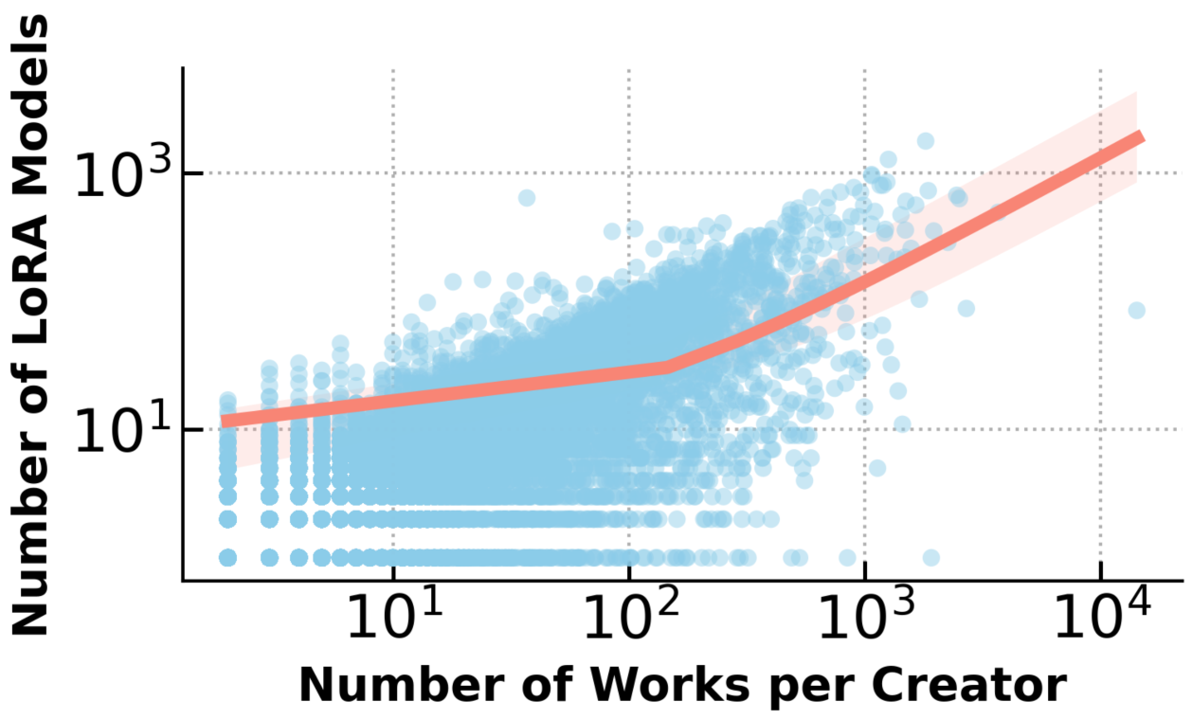}
    \vspace{-2ex}
    \caption{The correlation between the number of models and the number of works per creator.}
    \vspace{-1ex}
    \label{fig:correlation_work_model_count}
\end{figure}

Figure \ref{fig:correlation_work_model_count} shows the number of models a creator use \vs the number of their works.
Both base and LoRA models show a positive correlation: creators who produce more works tend to use a larger number of unique models. 
This is expected, since the number of unique models a creator uses is bounded by their total number of works.
However, the relationship is not straightforward. 
For a significant portion of creators, particularly those with a moderate number of works (\eg <200), the correlation is weak. As seen in the dense horizontal bands of points at the bottom of both plots, many creators in this bracket stick to a very limited repertoire of models, often using fewer than 5 base models. This behavior likely contributes largely to the overall market concentration, where a small subset of popular models is used heavily.
Conversely, the drive for diversity appears to be led by the most prolific creators (\eg > 500 works). These users experiment with a much wider array of both foundational base models and specialized LoRAs, suggesting that as creators become more experienced and invested in the ecosystem, they are more likely to explore and leverage its diversity.

These findings imply that many creators may not take advantage of the wide range of available models, perhaps due to the technical or cognitive overhead of discovery and experimentation. This points to a potential need for better tools to support creators. A model recommendation system, or even features for automatic model selection and synthesis, could help lower the barrier to entry for experimentation, enabling a broader user base to harness the full creative potential of the open-source ecosystem.

\section{RQ2: Model Life Cycle}
\label{sec:rq2}

This section further investigates the temporal dynamics of model usage. We examine the life cycle of generative models within the Pixiv community, tracking models from their release through their peak popularity to their eventual decline as newer alternatives take their place. Studying these life cycles will reveal how quickly creators adopt new tools or stick to old habits, providing insights that help developers and platforms better support their users.

\subsection{Temporal Analysis of Model Popularity}
\label{subsec:rq2:temporal_popularity}

\pb{Methodology.}
To begin our temporal analysis, we first define a normalized weekly popularity score for each model. This score is calculated by dividing the number of images a model produced in a given week by its maximum weekly output observed in our dataset. This normalization is essential, as it allows for a direct comparison of the life cycle between models of vastly different overall usage volumes.
The publication time is from the Civitai data as detailed in \S\ref{subsec:data:model}. 
For the models that are not found on Civitai, we estimate using the first-use time. To ensure that we get enough data for each model, we only include the models that are used at least 1000 times. 

\begin{figure}[t]
    \centering
    \includegraphics[width=\linewidth]{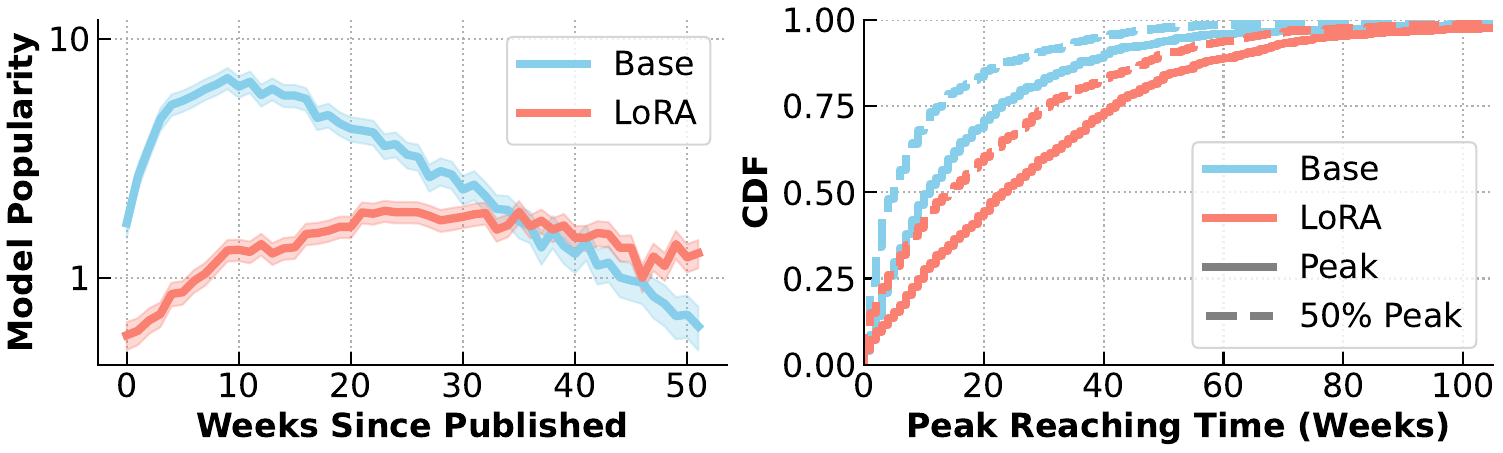}
    \vspace{-4.3ex}
    \caption{(a) Weekly popularity after model publication, averaged over models used for more than 1000 times; (b) CDF of the peak reaching time of the models.}
    \vspace{-2ex}
    \label{fig:rq2_popularity_vs_time}
\end{figure}

\pb{Result.}
Figure \ref{fig:rq2_popularity_vs_time}a presents the average popularity trajectory for base and LoRA models during the first year after their publication. 
We observe that base models exhibit a distinct ``rise and fall'' pattern in their life cycle. While they achieve their maximum usage quicker than LoRA models, their adoption is still a slow process. Popularity builds gradually over the first 10 weeks before reaching a peak and subsequently entering a steady, prolonged decline. 
Figure \ref{fig:rq2_popularity_vs_time}b, which plots the CDF of the time required for models to reach their highest popularity, further illustrates this sluggish adoption. It shows that it takes 5 weeks for 50\% of base models to reach half of their peak usage, and a 11 weeks to hit absolute peak usage. Following this peak, the near-linear downward slope on the logarithmic y-axis in Figure \ref{fig:rq2_popularity_vs_time}a indicates an exponential decay in usage, presumably as these models are superseded by newer alternatives.

In contrast, LoRA models follow a ``rise and sustain'' pattern. Their adoption is more gradual, with popularity climbing over approximately 20-25 weeks. Indeed, Figure \ref{fig:rq2_popularity_vs_time}b quantifies this slower pace: it takes 23 weeks for 50\% of LoRA models to reach their peak popularity, which is much slower than base models. However, after this initial growth, their usage does not decline but instead enters a stable plateau, suggesting they maintain their relevance and utility for a much longer period within the observed timeframe.

Overall, we observe clear but different life cycle patterns for base and LoRA models, which are aligned with their distinct roles. Base models are foundational and subject to rapid obsolescence, leading to a ``rise  and fall'' popularity curve. LoRAs, however, are specialized tools that fulfill a persistent, often personalized or customized need, resulting in a pattern where their utility endures over time. 
This underscores the contrast between the open generative AI ecosystem and close-source platforms like Midjourney --- it evolves through both the constant replacement of core technology and the steady accumulation of durable, specialized creative assets.
However, the lengthy time it takes for both model types to reach their peak usage raises a notable concern regarding the diffusion of innovation. Despite the availability of new tools, the community's actual transition to these upgrades remains slow, possibly due to the friction of manually updating established workflows and the challenge within a ``decentralized'' ecosystem to discover and agree that a model is superior. This highlights a pressing need to develop intelligent recommendation systems and automated workflow adaptation tools to streamline the discovery and integration of emerging models.

\subsection{Analysis of Model Version Update}
\label{subsec:rq2:version}

A model is rarely a static tool; it often evolves through successive versions released by its developers (\ie v1 -> v2 -> v3). Therefore, to better understand the observation that individual models may take time to gain popularity (\S\ref{subsec:rq2:temporal_popularity}), we now analyze the \textbf{version update} dynamics within model families. 

\begin{figure}[h!]
    \centering
    \includegraphics[width=\linewidth]{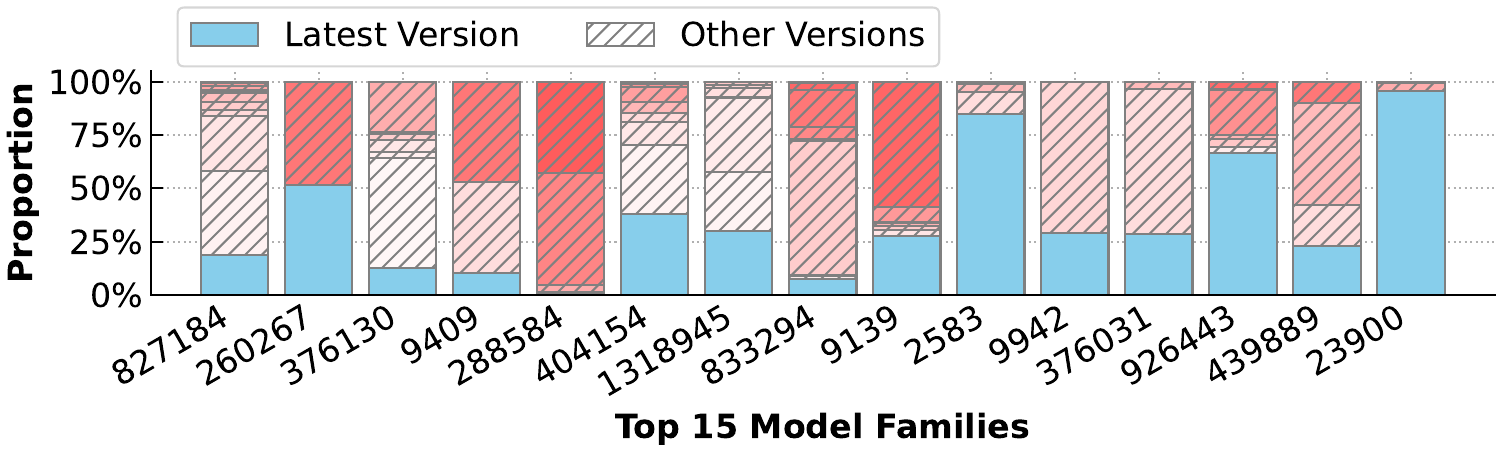}
    \vspace{-4.3ex}
    \caption{Breakdown of the version usage for the top-15 most used model families after the latest version's publication.}
    \vspace{-2ex}
    \label{fig:rq2_top15_version_proportion}
\end{figure}

\begin{figure}[h!]
    \centering
    \includegraphics[width=\linewidth]{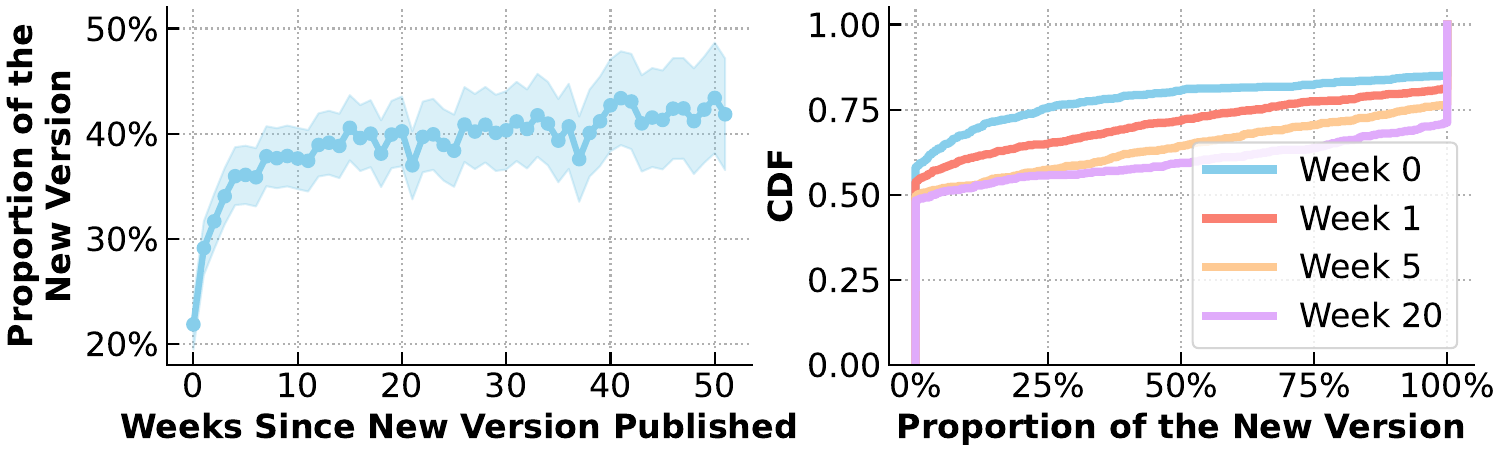}
    \vspace{-3ex}
    \caption{(a) Weekly proportion of the new version usage out of its model family after its publication; (b) CDF of the proportion of the new version usage out of its model family.}
    \label{fig:rq2_version_vs_time}
\end{figure}

\pb{Latest Version Adoption.}
We find that users exhibit significant inertia in adopting new model versions. Even for the most popular model families, older versions often retain a substantial (or even dominant) share of usage long after a new version is released. As illustrated in Figure \ref{fig:rq2_top15_version_proportion}, for 11 of the top 15 models, the latest version constitutes less than half of the total images created within that family.
To further understand the dynamics of this adoption process, we analyze the uptake of new versions over time where Figure \ref{fig:rq2_version_vs_time}a plots the result.
The adoption is a slow and gradual process. On average, a new version's usage share within its model family starts at about 20\% in its first week and quickly grows to around 40\% in 10 weeks. However, for the following 40 weeks, there is only a slight increase of less than 5\%.
Figure \ref{fig:rq2_version_vs_time}b shows the CDF of the proportion of the new version usage, which further underscores this slow uptake. It highlights the fact that, over 50\% of new versions see no usage, even after 20 weeks. However, it also reveals a polarized adoption pattern --- as of week 20, 80\% of models experience either no usage of the latest version or a complete adoption. Overall, we see a substantial and persistent share for older versions, which explains the slow peak-reaching speed and long life cycles of individual model versions observed in \S\ref{subsec:rq2:temporal_popularity}.

\begin{figure}[h!]
    \centering
    \includegraphics[width=\linewidth]{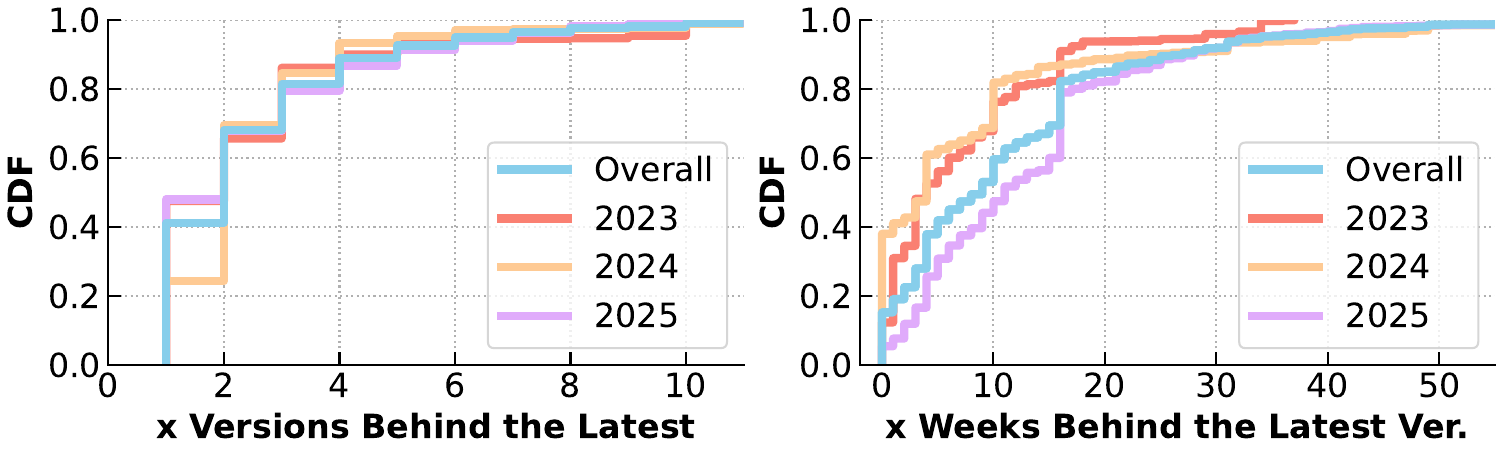}
    \vspace{-4ex}
    \caption{CDF of the images not using the latest version: (a) how many versions the model used is behind the latest; (b) how many weeks the model used is behind the latest.}
    \vspace{-1.5ex}
    \label{fig:rq2_version_behind_latest}
\end{figure}

\pb{Reasons behind the Inertia in Adopting New Versions.}
We now investigate the reasons behind the observed inertia in adopting new model versions. A straightforward explanation might be that creators are willing to update but are simply late to switch, rather than deliberately using old models. To test this, we measure how old the non-latest versions in use actually are. 
Figure \ref{fig:rq2_version_behind_latest}a shows the CDF of how many versions an image's model is behind the latest available version of its family, while Figure \ref{fig:rq2_version_behind_latest}b shows the CDF for how many weeks it is behind.

The data partly supports our hypothesis: old versions in use are typically not very old. Approximately 40\% of images using the non-latest model version are created with a model that is just one version behind the latest (Figure \ref{fig:rq2_version_behind_latest}a). 
The time lag analysis reveals a similar pattern: about 18\% of these images are created less than one week after a newer version was released, and this figure is 40\% within four weeks (Figure \ref{fig:rq2_version_behind_latest}b). 
Overall, this suggests that a large proportion of the observed inertia is not due to a preference for older models, but rather a natural delay in the  update cycle.

However, this delay in adoption does not fully explain the behavior of the remaining large proportion of users who are more than a few weeks or a single version behind. 
A crucial aspect of this ecosystem is that price and hardware requirements are generally not a factor, as version updates do not typically make a model larger or more resource-intensive. 
Thus, retaining an older version is likely driven by a combination of habit and creative preference. Creators may stick with the models and workflows they already know and, if the new version does not provide a significant improvement, they may see little incentive to disrupt a familiar process. Additionally, some creators may genuinely prefer the specific aesthetic qualities, style, or behavior of an older version that might have been altered or lost in a newer version. 
This explanation is also supported by an analysis of comments on Civitai, with detailed information provided in Appendix \S\ref{appendix:civitai_comment}.

This phenomenon of old version preference is not unique to this community; similar behavior is seen in other (generative AI) ecosystems, such as the ``keep4o'' movement for ChatGPT, where users preferred the GPT-4o version and when OpenAI replaced GPT-4o with GPT-5, it triggered an intensive user resistance movement \cite{lai2026please}. This highlights a significant strength of the open-source model community. Unlike in commercial environments where corporations can force updates and deprecate older versions, users here retain the freedom to choose their preferred tool. This autonomy ensures that a diverse range of model versions, each with its unique characteristics, can coexist and remain accessible, empowering artists with full control over their creative process.

\section{RQ3: Use \& Composition of LoRA Models}
\label{sec:rq3}

The ability for users to create, share, and combine lightweight LoRA adaptations with a larger base model is a hallmark of the open-source generative AI community.
Understanding these emergent workflows is crucial. It not only reveals the evolving artistic practices within the community but also provides vital feedback for model developers and platform designers on what users value and where current tools fall short.
Therefore, this section explores how creators employ and combine LoRAs, identifying the evolution of patterns and strategies and their effectiveness.

\subsection{LoRA Model Adoption}
\label{subsec:rq3:lora_adoption}
We start by examining the adoption trends of LoRA models to verify two aspects: \one Is the adoption of LoRAs prevalent, and how does it evolve over time? \two Does the use of LoRAs have a positive impact?

\begin{figure}[h!]
    \centering
    \includegraphics[width=\linewidth]{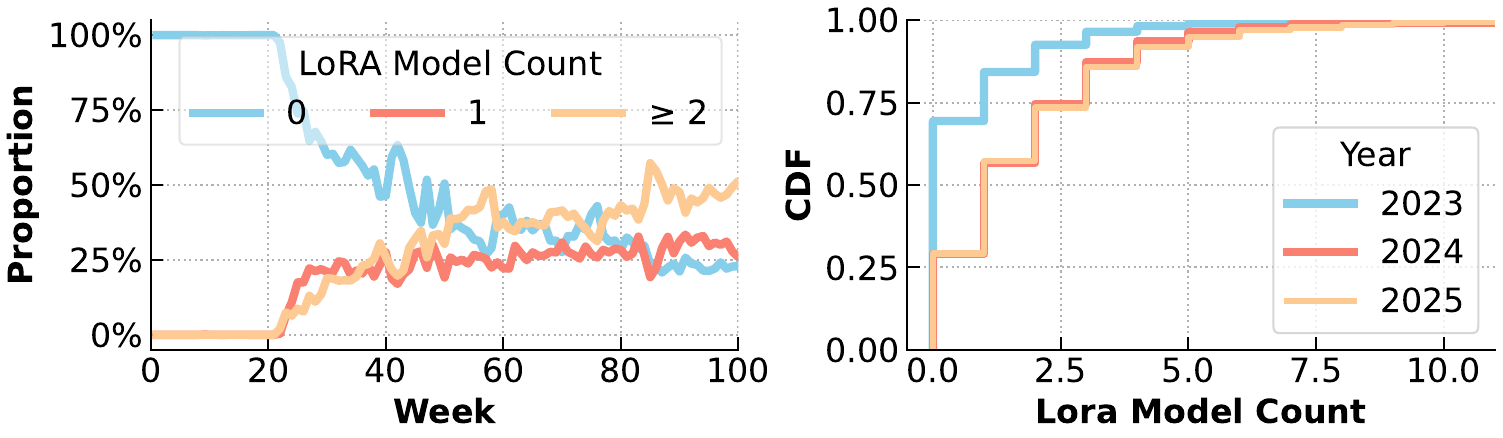}
    \vspace{-4ex}
    \caption{(a) Weekly proportion of images using LoRA models; (b) CDF of the number of LoRAs used by the images.}
    \vspace{-2ex}
    \label{fig:rq3_lora_count_vs_time}
\end{figure}

\pb{Adoption over Time.}
First, we look into the usage of LoRA models over time.
As shown in Figure \ref{fig:rq3_lora_count_vs_time}a, LoRA use has quickly evolved to a dominant practice since its emergence in Week 22. 
Initially, the use of a single LoRA became prevalent, but this was soon overtaken by the practice of using two or more LoRAs. By the end of our observation period, the ecosystem reached a new equilibrium where the majority of images (75\%) utilize at least one LoRA, with 50\% of them using two or more. This confirms the prevalence of LoRA usage, where creators move away from base models alone and towards increasingly complex workflows. 

Figure \ref{fig:rq3_lora_count_vs_time}b shows the number of LoRAs used by the images. The data shows that the majority of LoRA-enhanced workflows remain relatively simple. For instance, in 2024 and 2025, 49\% of all images used one or two LoRAs. This indicates that using a small number of LoRAs is the most common practice. However, the lengthening tail of the CDF curves for 2024 and 2025 reveals a growing trend of experimentation: a small but increasing number of users are stacking three or even more LoRAs, pushing the boundaries of customization. This suggests a dual pattern: widespread adoption of basic LoRA usage, alongside the emergence of a community engaged in more complex model composition, which motivates us to further explore how creators combine multiple LoRAs in \S\ref{subsec:rq3:lora_category}.

\begin{figure}[h!]
    \centering
    \includegraphics[width=\linewidth]{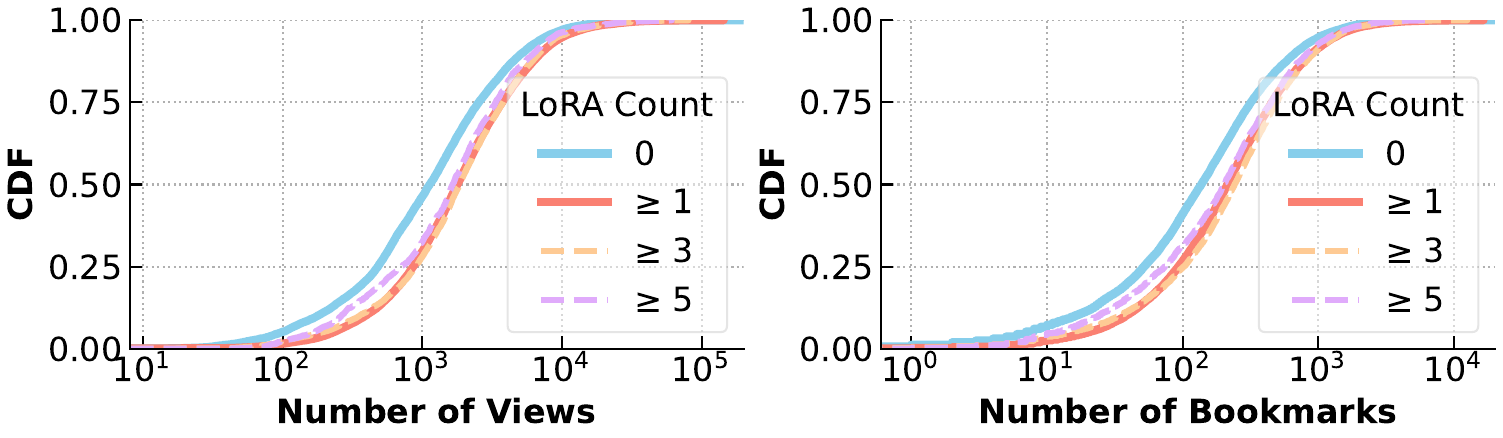}
    \vspace{-4ex}
    \caption{CDF of the number of the (a) views and (b) bookmarks received by the images uploaded during \texttt{2025-04}.}
    \vspace{-2ex}
    \label{fig:rq3_lora_count_vs_popularity}
\end{figure}

\pb{LoRA Adoption \vs Artwork Popularity.}
Having established the growing adoption of LoRAs, we now investigate whether the use of LoRAs have a positive impact. To do this, we examine the correlation between LoRA usage and artwork popularity, as measured by the number of views and bookmarks.
Figure \ref{fig:rq3_lora_count_vs_popularity} presents the CDF of these popularity metrics for artworks grouped by their LoRA count. 
Note, to enable a fair comparison, we investigate the artworks published in the same month, and here the figure plots the result for \texttt{2025-04} as a typical example (results for other months can be found in Figure \ref{fig:rq3_lora_count_vs_popularity_all} in supplementary material).

A clear pattern emerges: artworks created with at least one LoRA consistently outperform those created without any. This suggests that the customization and refinement enabled by LoRAs are correlated to an artwork's success.
However, the relationship between the number of LoRAs and popularity appears to be one of diminishing returns. The curves for using more than three or five LoRAs show only marginal differences. Notably, there are 5.6\% of the artworks generated with 5+ models, while in some regions, the ``$\ge 5$'' curve  is even slightly to the left, suggesting a potential negative effect.
This implies that while the act of using LoRAs can be beneficial, simply stacking more of them is not an effective strategy for maximizing an artwork's reach. The key benefit does not come from the sheer complexity of the model composition.

\subsection{LoRA Model Category \& Combination}
\label{subsec:rq3:lora_category}

\S\ref{subsec:rq3:lora_adoption} shows the prevalence of (multi-)LoRA workflows. However, simply counting the LoRAs used does not reveal the substance of this customization. 
To gain a deeper understanding, it is essential to investigate what kind of LoRAs are employed and uncover the primary intentions behind their use, \eg introducing specific content or refining aesthetics and workflows. Therefore, we examine the categories of LoRAs and explore how they are used and combined. 
We use the category from the Civitai data as described in \S\ref{subsec:data:model}.

\begin{figure}[h!]
    \centering
    \includegraphics[width=\linewidth]{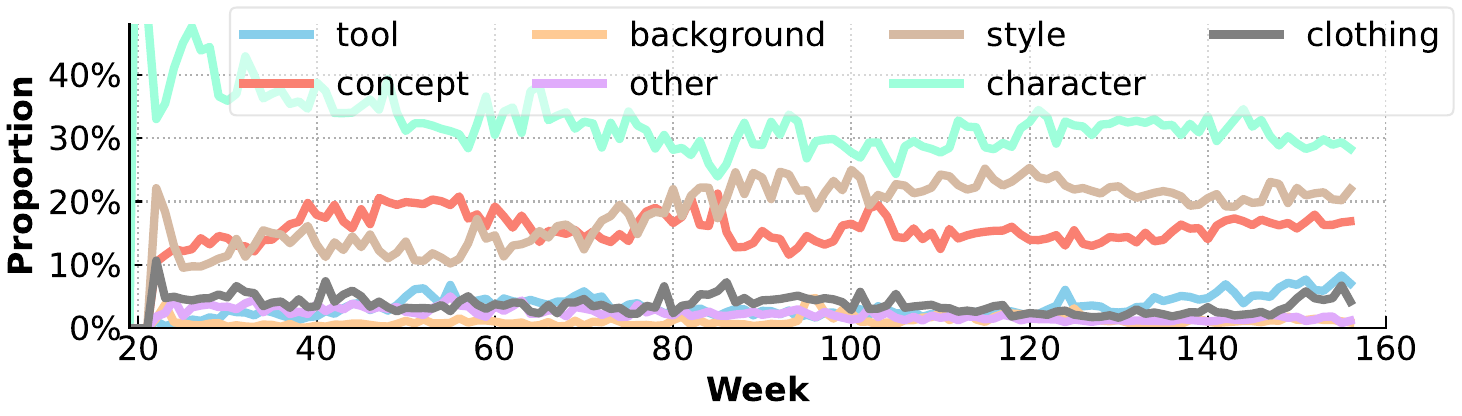}
    \vspace{-3.3ex}
    \caption{Weekly proportions of the LoRA categories.}
    \vspace{-2ex}
    \label{fig:rq3_lora_category_vs_time}
\end{figure}

\pb{LoRA Category Distribution.}
Figure \ref{fig:rq3_lora_category_vs_time} shows the usage proportions of the LoRA categories over time.  Throughout the observed period, character LoRAs consistently constitute the largest share of usage. However, their dominance is waning, with their proportion showing a gradual downward trend from a peak of over 40\% to a more stable level around 30\%.
In contrast, the use of style LoRAs exhibits a upward trajectory. Their share has grown steadily, eventually surpassing concept LoRAs to become the second most prominent type, accounting for over 20\% of usage by the end of the period. Meanwhile, concept LoRAs maintain a relatively stable and significant share, consistently hovering around the 15-20\% mark. 

The remaining categories represent a small fraction of total LoRA usage, each typically comprising less than 5\%, however, we also see a recent increase of clothing and tool, approaching 10\%. 
This evolving distribution suggests a shift in user focus, moving from an initial primary concern with rendering specific \textit{characters} towards the aesthetic \textit{style} and thematic elements (\textit{concept}) of their artworks, with the help of \textit{tools} for more more nuanced control over.

\begin{figure}[h!]
    \centering
    \includegraphics[width=\linewidth]{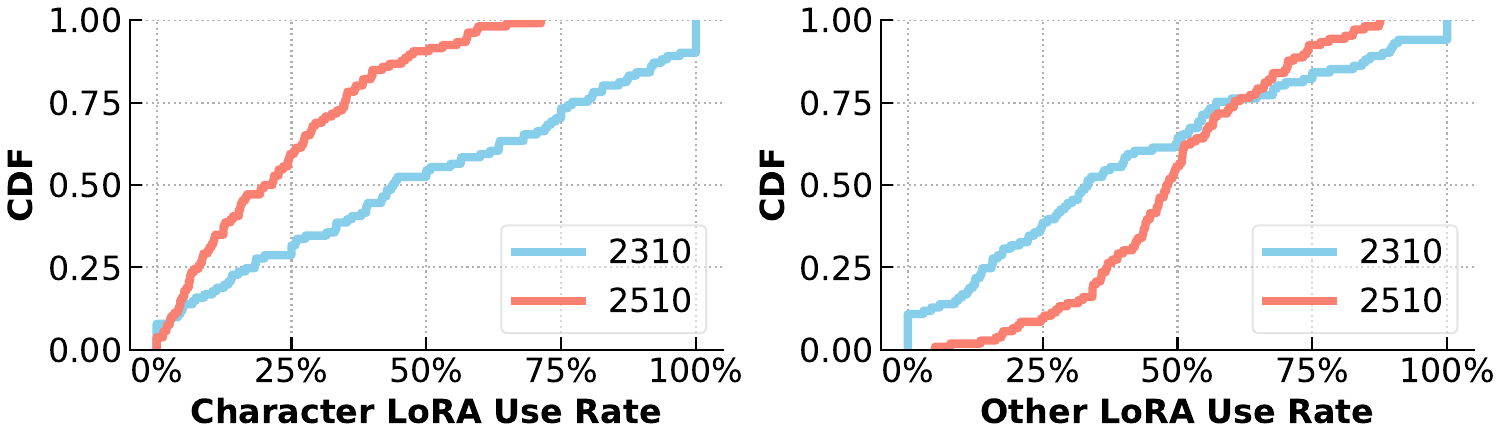}
    \vspace{-4ex}
    \caption{CDF of the (a) character LoRA use rate and (b) other LoRA use rate of the images depicting a selected character.}
    \vspace{-1ex}
    \label{fig:rq3_lora_use_character}
\end{figure}

\pb{Why the Decline of Character LoRAs.}
The change of usage proportions, particularly the relative decline of character-category LoRAs, warrants further investigation. 
By understanding what kind of models are increasingly in demand and why, creators can better focus their efforts and develop and share more effective models.
We conjecture that the reason is that as base models develop, they become easier to use and are often pre-trained with extensive knowledge of a wide range of characters; thus, the same character that needs a LoRA model before can be generated with a simple prompt.  For example, the most popular model family in our dataset, \textit{WAI-illustrious-SDXL}, explicitly notes in its newer versions that it supports over 4,400 characters with direct prompting \cite{wai_support_character}. This also aligns with our earlier finding that different base models fulfill distinct functional roles in terms of LoRA usage (\S\ref{subsec:rq3:lora_adoption}).
To empirically test this, we select a set of 108 popular characters (see Table \ref{tab:popular_char} in supplementary material for details) and track the rate at which they are generated with and without LoRAs. We compare two periods: \texttt{2023-10} (around Week 40) and \texttt{2025-10} (around Week 150). Figure \ref{fig:rq3_lora_use_character} presents the CDF of LoRA usage rates for these characters across the two periods, yielding strong evidence for our conjecture.

The left panel confirms a significant drop in reliance on character-specific LoRAs: the median character's LoRA usage rate plummeted from approximately 45\% in \texttt{2023-10} to just 22\% in \texttt{2025-10}. Conversely, the right panel shows that users are applying more non-character LoRAs, with the median usage rate for these types increasing from 30\% to 49\% over the same period. 
Taken together, these findings illustrate how improving base models reshape LoRA usage. 
Previously, creators understand the fact that LoRAs can conflict with each other, and our earlier findings also confirm that stacking too many LoRAs risks diminishing the popularity (\S\ref{subsec:rq3:lora_adoption}).
However, by relying on the latest base model to render the subject (in this case, the character), creators free up their limited ``LoRA budget'' for style and concept modifications, allowing them to safely apply aesthetic customizations without overloading the generation process and degrading image quality.
This shift represents a positive evolution, where highly specialized style and concept LoRAs are pushing the boundaries of artistic customization.
However, this trend also carries a negative implication: usage of base models may further concentrate on a few highly capable, dominant ones, which corresponds to the with the result in \S\ref{subsec:rq1:distribution}, Figure \ref{fig:rq1:entropy}. This centralizing effect risks reducing base model diversity and might stifle broader open-source model innovation in the long term.

\begin{figure}[h!]
    \centering
    \includegraphics[width=0.9\linewidth]{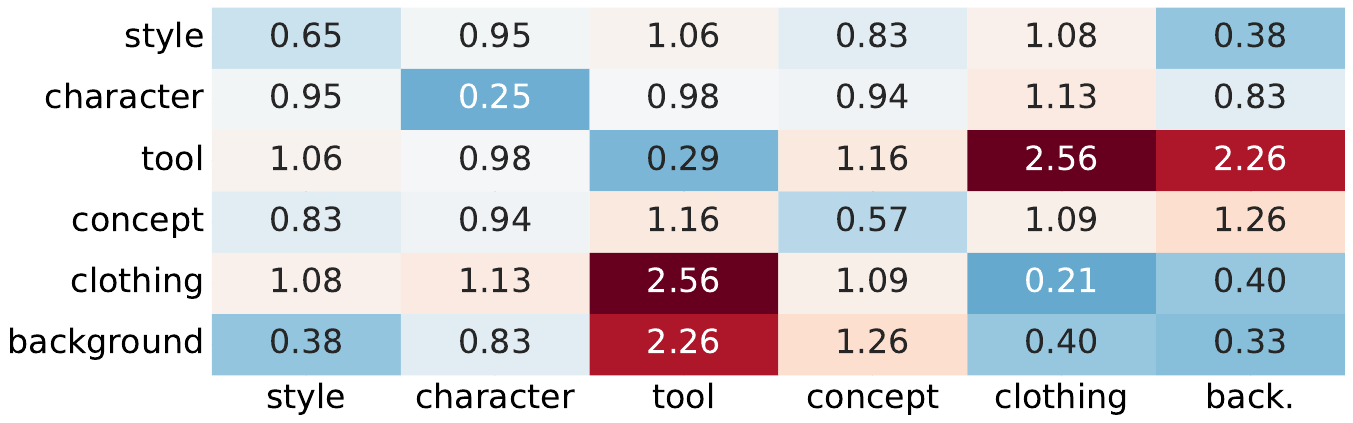}
    \vspace{-2ex}
    \caption{Co-occurrence ratio between two LoRA categories for images using two or more LoRAs as of \texttt{2025-Q4}.}
    \vspace{-2ex}
    \label{fig:rq3_lora_category_cooccurrence}
\end{figure}

\pb{LoRA Combination and Co-occurrence.}
Given that 50\% of the images are generated using two or more LoRAs, understanding the workflows involved in multi-LoRA setups is crucial. Therefore, we look into the combinations of LoRAs to identify which model categories naturally synergize and which are typically avoided. To achieve this, we examine the patterns of co-occurrence among different LoRA categories.
To quantify these patterns, we first established a baseline occurrence rate ($R_1$) for a given Category A. Next, we calculated the conditional occurrence rate ($R_2$) of Category A given the presence of Category B. The co-occurrence ratio between A and B was then defined as $R_2 / R_1$, where a value greater than 1.0 indicates that the categories are more likely to be paired together. The results are shown in Figure \ref{fig:rq3_lora_category_cooccurrence}.

We first see which model categories synergize effectively.
A clear pattern emerges for tool and clothing LoRAs: they exhibit a strong positive co-occurrence with nearly every other LoRA type, most notably with each other (2.56).
This suggests they are treated as utility modules that can be effectively added to a wide variety of workflows to refine specific aspects of an image. This versatility also provides an explanation for their gradually growing popularity (Figure \ref{fig:rq3_lora_category_vs_time}), suggesting that they are becoming a more integral part of the creator's toolkit. On the other hand, many pairs, such as style with character (0.95), have co-occurrence scores near 1.0. This neutrality suggests they are combined based on specific image requirements rather than being deliberately sought out or avoided. In contrast, creators actively avoid combining LoRAs of the same type, as shown by exceptionally low scores for character (0.25) and clothing (0.21) pairs. This is likely to prevent conflicting instructions and maintain a coherent artistic vision.


Overall, these patterns imply an evolution towards a sophisticated, component-based workflow where creators strategically layer specialized  LoRAs to refine their vision. This demonstrates a deep community understanding of model composition. However, this complexity may also form a barrier for new/amateur creators, which highlights a need for future tools that can help users navigate the usage of LoRAs and their compatibility and conflict.

\vspace{-1ex}
\section{Conclusion}

In this paper, we have presented a large-scale empirical study of the open-source generative AI ecosystem. By constructing and analyzing a novel dataset of millions of images from a major creative community, we systematically investigated how creators select, combine, and utilize a vast and diverse array of community-contributed models.
Our findings underscore the ecosystem's core strength: a highly modular and flexible environment that grants creators unparalleled agency over their artistic process. This paradigm enables sophisticated, customized workflows through the combination of foundational and specialized models, fostering a degree of artistic specialization not possible within  closed-source alternatives.

\pb{Recommendations.}
However, this richness also introduces challenges. 
Our analysis reveals a friction between the act of creation and the sharing and upgrading of new models. Reducing this friction by more tightly integrating creation tools with sharing platforms is crucial for accelerating innovation and lowering the barrier to entry for new model contributors.
Further, the sheer volume of options can overwhelm creators, and navigating the complex, ever-evolving landscape of models and techniques demands considerable expertise. To sustain the vitality of this open paradigm, future efforts should focus on developing intelligent tools and support systems that can help creators discover models, manage complex workflows, and fully harness the ecosystem's creative potential.

\bibliographystyle{ACM-Reference-Format}
\bibliography{sample-base}

\clearpage
\appendix
\section{Appendix}

\subsection{Analyzing the Comments on Civitai}
\label{appendix:civitai_comment}
To further explore why creators continue using older versions, we analyzed user comments on models hosted on Civitai. We randomly selected 1,000 comments from a total of 6,636, which represents a sample size of 15\%, focusing on the top 15 model families (as depicted in Figure \ref{fig:rq2_top15_version_proportion}). Through manual examination, we found that 11.2\% of the comments express a preference for older versions and/or dissatisfaction with the latest version. Most comments pertain to aesthetic styles or quality, such as: ``\textit{V10 feels like step back from V9. Too cartonish.}'' 
However, 14.3\% of the comments also highlight the LoRA compatibility issue: ``\textit{I think v9 is best for now. Many of characters Lora I use look weird after v10.}''

The LoRA compatibility issue indicates an extra friction; creators not only stick with familiar models and workflows but rather face substantial rework such as retraining LoRAs when migrating to a new version. Overall, this analysis confirms that creators can deliberately choose an older version because they do not perceive the latest release as a definitive improvement for their needs. The full list of these 112 comments is provided in Table \ref{tab:all_comments}.

\newpage

\subsection{List of Popular Characters}
Table \ref{tab:popular_char} lists the 108 popular characters used for the analysis in \S\ref{subsec:rq3:lora_category}. The list is based on the popular tags from the official Pixiv website.\footnote{\url{https://www.pixiv.net/tags}} The character is either featured as a popular tag or comes from a work/theme that is listed as a popular tag.

\begin{CJK}{UTF8}{min}
\begin{table}[h!]
    \centering
    \footnotesize
    \begin{tabular}{ll}
    \toprule 
        {博麗霊夢} & {霧雨魔理沙} \\
        {フランドール} & {レミリア} \\
        {十六夜咲夜} & {魂魄妖夢} \\
        {古明地こいし} & {古明地さとり} \\
        {射命丸文} & {アリス・マーガトロイド} \\
        {チルノ} & {藤原妹紅} \\
        {パチュリー} & {東風谷早苗} \\
        {八雲紫} & {アルトリア・ペンドラゴン} \\
        {マシュ・キリエライト} & {ジャンヌ・ダルク} \\
        {スカサハ} & {遠坂凛} \\
        {間桐桜} & {ネロ・クラウディウス} \\
        {沖田総司} & {モルガン} \\
        {メリュジーヌ} & {宮本武蔵} \\
        {玉藻の前} & {イリヤ} \\
        {ジャンヌ・ダルク〔オルタ〕} & {BB} \\
        {砂狼シロコ} & {一之瀬アスナ} \\
        {角楯カリン} & {早瀬ユウカ} \\
        {聖園ミカ} & {空崎ヒナ} \\
        {天童アリス} & {下江コハル} \\
        {陸八魔アル} & {小鳥遊ホシノ} \\
        {伊落マリー} & {才羽モモイ} \\
        {才羽ミドリ} & {秤アツコ} \\
        {錠前サオリ} & {天雨アコ} \\
        {花岡ユズ} & {生塩ノア} \\
        {銀鏡イオリ} & {美甘ネル} \\
        {島風} & {金剛} \\
        {鹿島} & {加賀} \\
        {赤城} & {榛名} \\
        {時雨} & {響} \\
        {天津風} & {浜風} \\
        {雷電将軍} & {ナヒーダ} \\
        {フリーナ} & {胡桃} \\
        {甘雨} & {八重神子} \\
        {神里綾華} & {宵宮} \\
        {申鶴} & {夜蘭} \\
        {エウルア} & {ニィロウ} \\
        {ナヴィア} & {三月なのか} \\
        {カフカ} & {花火} \\
        {ホタル} & {黄泉} \\
        {符玄} & {銀狼} \\
        {ブローニャ} & {停雲} \\
        {鏡流} & {ルアン・メェイ} \\
        {宝鐘マリン} & {兎田ぺこら} \\
        {星街すいせい} & {がうる・ぐら} \\
        {白上フブキ} & {湊あくあ} \\
        {壱百満天原サロメ} & {月ノ美兎} \\
        {星野アイ} & {星野ルビー} \\
        {有馬かな} & {フリーレン} \\
        {フェルン} & {ヨル・フォージャー} \\
        {アーニャ・フォージャー} & {喜多川海夢} \\
        {後藤ひとり} & {伊地知虹夏} \\
        {錦木千束} & {井ノ上たきな} \\
        {甘露寺蜜璃} & {胡蝶しのぶ} \\
        {初音ミク} & {2B} \\
    \bottomrule
    \end{tabular}
    \caption{Popular Characters.}
    \label{tab:popular_char}
\end{table}
\end{CJK}

\onecolumn
\begin{longtable}{ 
    >{\small\raggedleft\arraybackslash}p{0.05\textwidth} 
    >{\small\raggedright\arraybackslash}p{0.90\textwidth} 
}
\caption{Quotation of the comments.}
\label{tab:all_comments} \\

\toprule
\textbf{No.} & \textbf{Comment} \\
\midrule
\endfirsthead

\multicolumn{2}{c}{{\bfseries \tablename\ \thetable{} -- continued from previous page}} \\
\toprule
\textbf{No.} & \textbf{Comment} \\
\midrule
\endhead

\midrule
\multicolumn{2}{r}{{Continued on next page...}} \\
\endfoot

\bottomrule
\endlastfoot


1 & Not exactly impressed with 3.0, it's kind of more... sterile and flat. I hope they won't remove 1.3 and 2.2 \\
2 & V3.1 turns most of the characters into a childish, chibi style. I hope this can be fixed. It's really annoying. LoRAs barely work on the 3.1 version. It fails to maintain the quality of trained LoRA compared to the 3.0 version.I hope you can fix it soon, unless you'll have to wait for another sponsorship. \\
3 & Why Close Online Use.V2 is the best version of the pony anime style. \\
4 & I hope more focus can be put towards fixing the issues with vpred, epsilon is just ugly to see in comparison now  \\
5 & V4 didn't need as much of a prefix or suffix, whereas v777 requires at least a quality prefix for good output. \\
6 & v10.0 is kind of weird \\
7 & v15 is not really good but the V11 is the best model i used so far \\
8 & V2 really doesn't play nice with AlignYourSteps schedulers. Not a big deal, just making a note. \\
9 & Sticking with V3 as V4 somehow reduced fidelity, style and flexibility for me. I'm using like 10 loras and long prompts in ComyUI, got a perfect Mix with V3. With V4 i couldn't find it yet.If you wonder what style I'm talking abaout you can see it in my images they're all V3.In addition I use style Loras: Kenva 0.5  ; Vixons Gothic Neon1 -  0.3 ; Concept Art Twilight 0.5  \\
10 & is just me or do 5.0 generate just color blurred picture some time when add in some extra prompts ?P.s i use fooocus (with 4060ti) \\
11 & I think v9 is best for now. Many of characters Lora I use look weird after v10. \\
12 & Why does the latest version 15 often create green or pink artifacts in genital area?It's like it's trying to censor.Any suggestions? \\
13 & \zh{最新的1.0版本测试下来感觉在各种概念上有了强化，但有个比较明显的问题，人物的大小腿比例很不正常，有的时候还容易扭曲，脚的崩坏率也比0.75版本更高。。。希望作者可以想办法修复下~} \\
14 & How to get dark skin girls, normal skin guys, V10 version is hard to get \\
15 & Sometimes the eyes get messed up for me. They are distorted and have multiple colors.I'm using Euler ancestral, 20 steps, 5 CFG in Comfyui. It happens with long prompts and short prompts. Any idea what I'm doing wrong? I didn't have this issue with V3. \\
16 & I was really looking forward to 15, so I was VERY disappointed with what it was.  The anatomy has degraded greatly, rolling back almost to the level of Pony6. I don't understand how it was possible to allow/not notice such blatant miscalculations during the checkpoint testing.  I've been out of the habit of ``necromorphism'' (a reference to the distorted forms of Dead Space monsters) of Pony/SDXL for almost a year now, so seeing this again was almost a shock.  Moreover, the mutations happen in fairly simple scenes and compositions that the previous versions had no problems with. I'm going back to 14. For me, the 15th is now the worst of all the existing iterations of WAI.  \\
17 & Great model, but I think it started getting worse after v7. Not sure if updates are really needed. \\
18 & C'mon make 14 2.4D nsfw version available for generation, it was one of the best illustrious models for onsite generation \\
19 & 16 is out great stuff!Edit: After testing this with many LoRa Models it seems hit or miss with me some characters look better then others for example this Checkpoint works great with all the W.I.T.C.H. LoRa Models I use, but with some others like characters from Bleach it not that accurate, I'll probably keep using 15 for most LoRa Models. \\
20 & it makes images that look very simple 2d. Like there is no VAE. I still like the mature version the best \\
21 & \zh{v9比以前的版本对比度低了} \\
22 & in 0.75, it applies lora styles perfectly but in 1.0 the style loras that I have trained are not being applied consistently to the images. Not sure if it is because the style loras needs to be retrained on this specific model for the style to be applied or not. \\
23 & V15 seems more homogeneous in style than V14. For matching specific source styles I have had more success with V14. In V15 I have also experienced the finger issues  that others have pointed out. \\
24 & Anyone find a replacement model for this? Images now coming out overly saturated and cartoonish, thanks \\
25 & I went from v11 to v13 and images lost the specular highlighting, so the colors look flat.  Is there a way to get that back?  I tried adding “specular highlight” as a tag, but that does nothing.  \\
26 & In version 2.1, most schedulers will not function properly, resulting in images that are not fully denoised. \\
27 & \zh{v13人物皮肤容易偏冷色暗色，配上高反光容易有塑料感，没有V12那种柔和的肉感；V13 相比V12更容易调出御姐脸型，这个好评} \\
28 & V10 feels like step back from V9. Too cartonish \\
29 & ... v16 seems nasty. Better posing, but worse adherence and a really messy look. It's like a 1.5 model. \\
30 & always getting big gigantic head on v4, anyone have a solution to avoid this? \\
31 & IF YOU GUYS WANT 1.0 BACK YOU HAVE TO BID FOR IT.  IT'S NOT UP TO THE CREATOR \\
32 & is v1.2 not able to generate in-site anymore? i was using it before and it was the best but now only v1.3 is available when i pick Hassaku as checkpoint. and im quite unsatisfied when i generate with my go-to lora preferences. (details, face, style etc.) or is it sth that i don't know? \\
33 & I use V-Pred 0.5 with some prompts but generate some bad picture, which not happen when I use E-pred. Its usage changed? \zh{(我使用V-Pred 0.5，和之前 E-pred 1.0 一样的prompt，但生成的图片非常奇怪，只是一些奇怪的色块，是使用方式变了吗？)} \\
34 & v11 por que lo quitaste f................. \\
35 & shame need to wait 7 days for test it my self then i read the discription of 2.2 and i think wait a minute hands worse sometimes darker? what?  tested 2.1 + 2.1fix i am not sure what it fixes exactly out of hands but 2.1 has on resolution 1024x1024 so much problems that i get cropped images like focus on the body and it does what it want if u write from side or something sometimes i get a img from behind and if i write not sky or clouds and only city the whole city is bright like alot of fog. so yeaah i apreciate the time that u spend in but i like the adult style taste of 2.1 and the drawing but what i hate on this is getting weird poses or cropped images ( the fog is ok if u write more stuff for the background )  if u think to work again on a 2.2 fix or maybe a better 2.1 then keep the adult style and watch on resolution 1024x1024 that u dont get cropped images , maybe it was a incomatible model that u merged in ur trained model so far i readed first the whole stuff and then i tested it :/  then i need to wait for a version thats like 2.1 drawing but not with the problems maybe it happens or maybe not i dont know its ur choice as creator but wish u good luck for the next versions that it works better greetings \\
36 & \zh{在提示词一样的情况下，为什么我用vpred版总是会出现人体结构和透视不是很准确的情况……有没有人懂怎么解决} \\
37 & v9 is still the goat V10 and v11 are extras(my opinion) \\
38 & V15 is pretty bad, It's still create good design. So V14 is the best \\
39 & I prefer 0.2 to 0.2M0.2 is more of a 3D anime style with higher quality and follows prompts better. Occasionally it turns out a low quality dud. Do not use ``masterpiece, worst quality'' etc. prompts with this.0.2M is consistent in a more drawn style. Must use ``masterpiece, worst quality'' prompts which can result in less prompt cohesion. \\
40 & I'm seeing watermarks and patreon logos in showcase images. This is not a good sign, and as some people have noticed, the overall quality is degrading with these recent versions.Please don't push version changes just for the sake of it, if they aren't upgrades. \\
41 & I just want to mention that with version 6, the eyes and fingers are of very poor quality; in version 5, using the same prompts, I get better results. \\
42 & Very bad to train LoRAs. 1.0 is ultra over-contrasted.You can see grid comparison:LoRAs trained on 1.0, 0.75 noobai and Illustrious (All used on NoobAI 1.0).LoRA which is trained on 1.0 is over-contrasted all time (even with artist style, which is not oversaturated).Illustrious LoRA forgets some details of character because between Illustrious and NoobAI a gap of 12 million art - It is not the same like PonyV6 and Autismmix where PonyV6 had shit over-contrasted style but autismmix (or style loras from pony page) fixed everything with small adding of style loras (obviously not with 12 million arts, lol).So Idk which is the best option to train loras now...I think I try 0.75. Maybe 1.0 loras will be good on future mixes and other finetunes of this models.P.S. Actually model is not for ``noobs'' because it doesn't support natural language (pony partially supports it), so danbooru tags only. And you must to know artists styles and good weights for them or style will be shit all time. \\
43 & I have tried V4 a bit and honestly, I am glad nothing drastic change (I welcome good change but sometimes it breaks my entire set up and redoing it can take days...)Some suggestions are the usual small anatomical hand fixes for gloves (or anything covering them, high chance of it being wonky in my gens, might be how niche it is tho) and usual general fixes that I wished came to SDXL based models from Flux Dev (the small details and my first impressions of it when I used generic prompts was something I never forget)But, one thing I seriously didn't like is the odd switch of the face returning to V1 but not the rest of the body, idk who suggested it but I BEG of you to change it back to what it was before as I didn't see any problems with it...I do like the better poses though and quality seems to be up? Sadly unlike back during V2 and 2.5 I'll be rolling back, as the face is REALLY off putting \\
44 & 5 changed too much.. everything seems to be flatter, less detailed and more cartoonish :/ \\
45 & I don't know whether is it just me or somehow I'm not able to use charecter LoRAs like the way I could use them in v8. I've having that problem with v10 and v11 (I've not tried the v9). \\
46 & From my testing v12 and v14 are very similar. I personally prefer v12, but both are good.  \\
47 & V2 siempre será la mejor \\
48 & This is still one of my favourite illustrious models, but it feels like the prompt-adherence has dropped a bit in the latest version. That's what i come to illustrious for, so this may become an impediment... \\
49 & Please add 1.0 back again! \\
50 & HANDS AND FINGERS ARE SOMEHOW WORST THAN IN AOM2 \\
51 & I don't know what you're doing, but after version 1.3, your checkpoint has gotten worse and worse.The checkpoints no longer produce proper anime-style images; everything is just weirdly semi-realistic.The checkpoints constantly add things that aren't listed in the prompt at all, like random objects, accessories, or entire people.And additional limbs or completely deformed limbs appear more frequently. \\
52 & Doesn't seem to generate loli as good as before and im having a lot of hand and artifact problem  \\
53 & \zh{v11版本感觉对画师tag压制很严重，适合不带画师的裸出图吧，在画师tag的情况下更喜欢v9，风格更明显，至于v10感觉在两者之间，感觉v9的基础上只需要增加lora的适配性，和新角色支持，这样出来的v9加强版比目前的v11,v10要好得多。} \\
54 & somehow in aom3 hands seems to be worst than aom2 i think \\
55 & Ver. 15 not bad, but i prefer ver. 14-12  \\
56 & After playing with 9R, I feel it's a hit or miss, it has better backgrounds, when it wants to make them, because for some reason does a of single-color backgrounds or it add a color filter to the full image. Red, Pink and Blue been some examples of filter I got into the image, I have tried changing prompts and CFG, but I haven't been able to pinpoint the cause. But I would get 1 or 3 out of 10 generations with what i wanted without issues, but the rest either not what I wanted or filters. It's odd, it excels in some things, but it suffers in others now where it didn't used to. I did notice it has better knowledge of some concepts now, which is good. Also, the photo database is mostly Asian, which I understand why, but variation would be good. Either way, I will keep testing it, but it's taking me longer to create what I envision in 9R compared to the previous version. \\
57 & In v13 I have artifacts with some models, as well as the image is not always clear. I compared it to version v12 and it works much better! I usually always use the latest version, but here it was so bad that I had to go back..... \\
58 & El nuevo modleo tiene problemas con las manos, el 14 es mejor \\
59 & V2 !!! V2 BEST FOR PONY FROM MLP !!! \\
60 & well, time to go back to 0.5. here's hoping they'll eventually update illustrious past 0.1 \\
61 & I've noticed that there seems to be a decline in image aesthetic, especially color, as this model continues to be trained. It appeared to have leveled off with 0.5, but the issue continues to be pronounced in 0.75. I implore the team to take a step back and make sure everything is working as intended. \\
62 & I tuned a few characters specifically for version 15, but it had overexposure issues (images were too bright) compared to 14. Hoping 16 fixes that.Quick question: I'm using 1152x896 resolution with 2x upscale, but the author recommends 1.5. Is that a big deal? \\
63 & can't do dark setting without throwing a flashbang at the chars. Bad prompt adherence for more complicated stuff, probably due do the dataset. using v12. maybe v11 is better. \\
64 & Well, the ship is sinking, so it's time to return to local generation. That brings us back to the big question, guys: V14 or V11? Do we go with V14 as it's the most current model (not counting V15), or do we pick V11, since it's often praised in the comments for being a solid model with great compatibility for many LoRAs? \\
65 & V8 is great.The V9 version has poor support for Hyper-SDXL, resulting in significant noise. I hope that the next version will restore good support for Hyper-SDXL. \\
66 & i think finger problem appears in v10 \\
67 & v12 somehow has a more blunt style with less details I would say (even with detail loras) compared to v11 \\
68 & V4 it's too dark \\
69 & am i the only one having face distortion issues with this new v11 update, specifically in the eyes? every model i use with this is having this problem. i tried using neg prompt tags like “bad anatomy” and “bad eyes” and it still doesn’t fix the problem. does anybody have any suggestions on things i could do differently? \\
70 & Comparing to 2.1fix, it seems I have more extra limbs/heads issue with 2.2 after testing for a few days. The same prompts often do not result in the extra limbs in 2.1f comparing to 2.2. \\
71 & vpred version seems degraded A LOT when generating kemono/furry subjuect, why happens this when you guys explicit say this model has a lot e621 training data? \\
72 & I'm still using v14. please more reviews for v15. lol \\
73 & Bring back V2 \\
74 & v13 not compatible for crotch pov prompt \\
75 & This version is probably better for inpainting but it's an absolute bullet to the brain for intricate artstyle mixes compared to 0.5 \\
76 & Why does the 1.3b version of the model generate images normally, but versions 2 and 2.1 produce only blurry noise, even when only the model is changed and all other settings remain identical? I’m not sure if I need to update anything, download additional plugins, or make any adjustments here. \\
77 & 4.0 feels weird to me. Poses and linework both seem over-exaggerated, and not in an appealing way, and skin is ridiculously shiny, it's like everyone is fully oiled up for some reason. Illustrious 2.0 or 2.5 feels like the sweet spot for me, I'm not sure what happened in 3.0 and especially 4.0, but man do I not like it. \\
78 & In my opinion, version 2.2 is the best \\
79 & I'm not sure, but 1.5 felt like it has more creativity/variety with a tradeoff with more body part merging/confusion between 2 subjects compared to 1.4 if you know what I mean. \\
80 & Not a fan of the new eyes version pls go back to previous 1.3 version \\
81 & I feel like 3.0 is still better with cleaner and clearer lines... \\
82 & I trained 3 lora with v10 and find it can hardly fit the face detail,while I didn't find this problem with v9,hope it can be fixed \\
83 & Hands are better now, but colors have become paler. Backgrounds and environment contain fewer interesting details. I like v1 than v2. \\
84 & V4 wasn't needed.  Go back to V2 and v3 \\
85 & RIP  v 11 \\
86 & Really like the model, however look like my lora is ``broken / noise'' when using v9, on v8 and v7 is still good. \\
87 & Yeah, I'm just gonna stick with V14, thanks. It just works better for my style \\
88 & Still love v1 illustrious, so underrated! \\
89 & v3 and v4 are clearly the best in general terms. v3 is the perfect midpoint between v4 and v6. V6 is a little bit more sharp and v4 has less contrast and overall a creamier look.v5 is complete trash ngl.Keep up the good work tho, creator! \\
90 & Quality output! I didn't do a huge amount of testing, but I had to correct minor color details twice. A tattoo color that didn't match the overall coloring and stood strangely out. Green eyeshadow where V14 chose a pleasing pink one. But that is my personal taste. Thank you for your work! \\
91 & v9 seems to have better lora compatibility and somewhat better backgrounds, but much less detail, larger chibi heads/eyes and overall leaning more towards such proportions compared to previous versions. Not that the previous ones had a decent maturebody basis, but this one even more so.Overall v7 is still the cleanest and can do great images without hires, but doesn't have decent lora support and is inflexible to style changes. \\
92 & bring back ver 2 \\
93 & Feels like v12 has LESS details and is more cartoonish than V11 \\
94 & Note to 0.4: Model is in a weird state, where it goes more and more to a 2d look now. Looks overall weird sometimes and have a bit of a ``blurry'' effect because of it. Next update is not in a week but in 1,5-2 weeks. \\
95 & I'm clearly not the only one who is bummed by the loss of v2, so I just posted a small visual comparison of v2, v3 and v4 and why v2 worked better for ME personally. I'm no expert and I just do image generation for my own entertainment so I don't really have much to add to the conversation, but I still felt it can be worth to give my 2 cents. Visually, at least. :) \\
96 & It's now actually and visibly worse than 1.3B. Worse details, more boring compositions. \\
97 & V14 was great. V15 produces a lot of images with 6 fingers or distorted mouths. I can't recommend V15 to anyone. \\
98 & I think the newest version makes heads look too immature. I feel like they looked better in v11. \\
99 & Lora compatibility seems to have decreased? Some LORAs I use cause more distortion than expected. I need to use less LORA and at lower strength. I can get good results without LORA or with LORAs with high compatibility. \\
100 & One Obsession is my favorite but I was getting much better results with v17 and v18. \\
101 & GO BACK 2.2, PLEASE MAKE IT AVAILABLE AGAIN! \\
102 & 5 is a downgrade, bring 4 back \\
103 & I agree with other commenters, 2.0 feels like a downgrade from 1.6.  \\
104 & Hi! Love your work, it's simply amazing.So I've been recently experimenting with mixes and merges, and I tried my hand at adapting the already existing stuff built on Anything v3 (basically everything at this point) to Anything v4, a different model which I suspect is wildly different to v3 when it comes to weights, because every mix I try to make using v4 comes out deformed and nowhere near the polish of their v3 counterparts - botched anatomy, far too much realism, the works. Even the composition differs a lot. My only successful model was when I MBW'd v4 and Basil after about five or so fiddlings with weights, and since then I was unable to add any other model to it, since any merge basically wrecks it.Do you have any advice on how to make v4 viable? I'm not sure if converting v3 models to v4 by adding difference (v4 - v3) with M=1 to existing v3 models is a viable strategy, but I could try that and see if it sticks. Thanks in advance! \\
105 & Why does this version take roughly 10 times longer to render an image compared to the previous version? Im not questioning the quality, but chances are I get better results from making 10 pictures from the v11 compared to a single one from v12. \\
106 & \begin{CJK}{UTF8}{min}私はv9以前に戻ります　V10以降手首とかの崩れ発生率が大幅に上がっていると感じますここまで早い速度で新しいVerを出さなくてもV8あたりで完成されてるmodelだと思っています\end{CJK} \\
107 & IMO, the ILV40 and ILV50 versions are almost identical in terms of image content. Additionally, the ILV50 provides images with a halo of fog. Weird. In my opinion, the ILV40 is better. \\
108 & This new one is awful, 14 was great but v9 remains mvp \\
109 & I can use lora to make faces more mature in V11-12, but V13-14 it disregards 90\% of the lora even if i set to 1, it cannot generate mature faces.  \\
110 & IMO v11 is the best, still using it a lot. The newer versions produce worse results for me. \\
111 & I really don't know how I feel about you're newest 18 model.....It seems extremely unstable at higher resolutions. I will do one more day of testing... \\
112 & Can you imagine ``cute girl on the balcony of a futuristic castle''? Once in beta version it was real girl in great inviroment. But what we see now?  \\

\end{longtable}

\begin{figure*}
    \centering
    \includegraphics[width=0.95\textwidth]{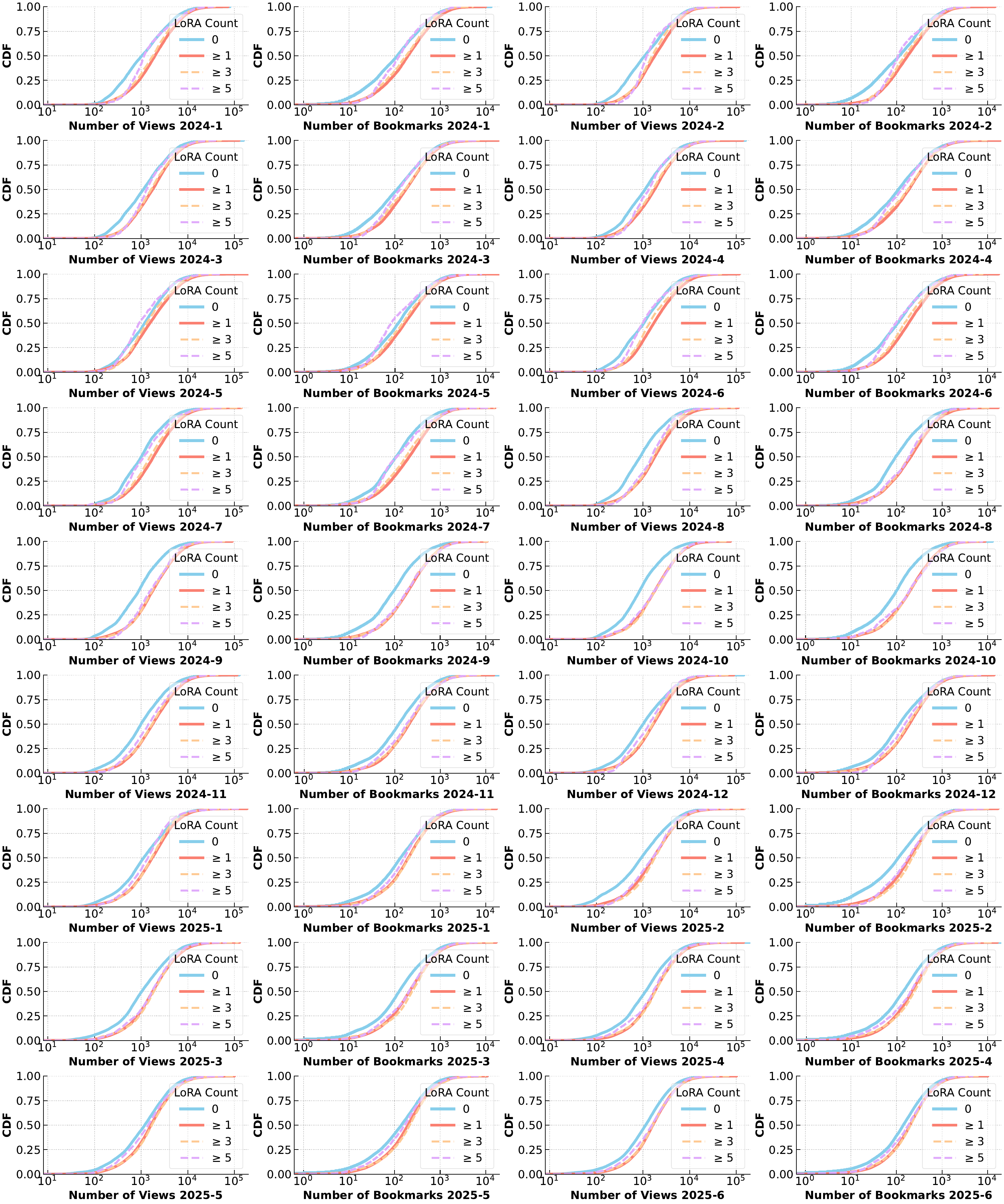}
    \caption{Monthly CDF of the number of the views and bookmarks received by the images.}
    \label{fig:rq3_lora_count_vs_popularity_all}
\end{figure*}

\end{document}